\documentclass[twocolumn,superscriptaddress,floatfix,preprintnumbers, nofootinbib]{revtex4-2} 
\pdfoutput=1
\usepackage[colorlinks=true,breaklinks=true]{hyperref}
\usepackage[normalem]{ulem}
\usepackage[utf8]{inputenc}
\hypersetup{allcolors=[rgb]{0.0 0.0 0.6},linkcolor=[rgb]{0.75 0.05 0.05}}
\usepackage{amsmath,amssymb}
\usepackage{epsfig}  
\usepackage{graphicx}   
\usepackage{slashed}       
\usepackage{tikz}
\usepackage{color}
\usepackage{url}
\usepackage{color}
\usepackage{multirow}
\usepackage{comment}
\usepackage{amssymb}
\usepackage{bm}
\usepackage{makecell}

\newcommand{\fermi}{\textit{Fermi}}
\newcommand{\sky}{\texttt{skyFACT}}
\newcommand{\onep}{\texttt{1pPDF}}
\newcommand{\sigmav}{$\langle \sigma v \rangle$}
\newcommand{\dnds}{$\mathrm{d}N/\mathrm{d}S$}

\allowdisplaybreaks

\bibpunct{[}{]}{,}{n}{}{,}

\begin{document}

\preprint{LAPTH-044/25}

\title{Stellar-like Galactic center excess challenges particle dark matter}

\author{Silvia Manconi}
\email{manconi@lpthe.jussieu.fr}
\affiliation{Sorbonne Universit\'e \& Laboratoire de Physique Th\'eorique et Hautes \'Energies (LPTHE),
CNRS, 4 Place Jussieu, Paris, France}
\affiliation{LAPTh, CNRS, F-74000 Annecy, France}

\author{Christopher Eckner}
\email{ceckner@ung.si}
\affiliation{Univ. of Nova Gorica, Center for Astrophysics and Cosmology, Vipavska 11, SI-5000 Nova Gorica, Slovenia}

\author{Francesca Calore}
\email{calore@lapth.cnrs.fr}
\affiliation{LAPTh, CNRS, F-74000 Annecy, France}

\author{Fiorenza Donato}
\email{fiorenza.donato@unito.it}
\affiliation{Department of Physics, University of Torino, via P. Giuria, 1, 10125 Torino, Italy}
\affiliation{Istituto Nazionale di Fisica Nucleare, via P. Giuria, 1, 10125 Torino, Italy}

\smallskip

\begin{abstract}
The Galactic Center (GC) is potentially hosting the largest indirect signal from particle dark matter (DM), which in many well-motivated models would produce gamma rays as their final states. However, this region has often been dismissed for DM studies because of the evidence
for an unexpected gamma-ray component over astrophysical backgrounds at GeV energies, firstly discovered in the data of the \textit{Fermi} Large Area Telescope (LAT), the so-called Galactic Center Excess (GCE). 
While this was initially considered to hint at GeV thermal relics, recent work supports a GCE interpretation  in terms of a stellar population of millisecond pulsar-like sources in the Galactic bulge. 
Building on this preference, we re-evaluate the GC as a powerful target for indirect DM searches via gamma rays. 
This is achieved by combining adaptive template fitting and photon-count statistical methods to assess the role of sub-threshold point sources in the observed \textit{Fermi}-LAT gamma-ray counts, while minimizing the mismodeling of Galactic diffuse emission backgrounds.
In a fully self-consistent way, the gamma-ray data are fitted with a mixed
model comprising a DM signal and a stellar bulge, both potentially contributing to the  GCE.
The space left for signals from weak-scale DM particle annihilations is quantified by extracting 95\% C.L. upper limits on the annihilation cross section, which, depending on the DM density profile, result in stringent limits for masses $\lesssim  300$ GeV. 
The robustness of our results is supported by tests on simulated data.

\end{abstract}

\maketitle

\section{Introduction}
\label{sec:introduction}
The indirect detection of dark matter (DM) particles through photons from pair-annihilation events is more promising in those sites where over-densities are predicted, such as the centers of DM halos, or where the signal-to-background ratio is expected to be particularly favorable, such as in dwarf spheroidal galaxies. 
The interest in the center of our Galaxy has, for a long time been faded due to the complexity of modeling the emitting sources along the line of sight, and especially in the Galactic center (GC) region. 
The Galactic center excess (GCE) measured by the {\em Fermi}-Large Area Telescope (LAT) \cite{Goodenough:2009gk,Vitale:2009hr,Hooper:2010mq,Hooper:2011ti,Abazajian:2012pn,Gordon:2013vta,Zhou:2014lva,Calore:2014xka,Daylan:2014rsa,TheFermi-LAT:2015kwa,DiMauro:2021raz,Murgia:2020dzu} is still lacking a clear explanation. These photons, which peak at a few GeV, are detected in excess of the diffuse Galactic emission, ascribable to known sources for energetic particles, which in turn interact with the interstellar medium and radiation fields. 
The interpretation of the GCE as due to particle DM annihilating into gamma rays is paralleled by the possibility of a population of dim point sources, such as millisecond pulsars (MSPs) \cite{Abazajian:2010zy}, emitting gamma rays below threshold. Indeed, when using a number of state-of-the-art, complementary techniques, the morphology of the excess turns out to be better described by a stellar population in the Galactic bulge than a Weakly Interacting Massive particles (WIMP) DM-inspired template peaked at the GC \cite{Bartels:2017vsx, Macias:2016nev, Macias:2019omb,Calore:2021jvg,Manconi:2024tgh,Song:2024iup}, typically assumed to be spherically symmetric. 
Specifically, the works reported in \cite{Pohl:2022nnd, Song:2024iup} performed a thorough, end-to-end comparison of analysis pipelines, masking strategies, diffuse-emission prescriptions, and available GCE spatial templates, explicitly in relation to claims favoring a more spherical morphology such as \cite{DiMauro:2021raz, Cholis:2021rpp, McDermott:2022zmq}. They find that the apparent preference for a spherical (DM-like) profile  can arise for particular combinations of comparatively rigid diffuse backgrounds and aggressive masking choices, and can be further impacted by fit convergence and by the specific implementation of the bulge template. However, once the astrophysical foreground is controlled with more elaborate techniques, e.g.~allowing greater freedom in the diffuse components (by ring-based descriptions and/or template modulation) and testing updated bulge morphologies (including the ``Coleman'' boxy bulge of \cite{Coleman:2019kax}), the data consistently favor a stellar-bulge-correlated excess over a purely spherical template, with spherical components becoming at most competitive in restricted corners of model space (see also \cite{Zhong:2024vyi}).
Besides, state-of-the-art magneto-hydrodynamical cosmological  simulations indicate a non-spherically symmetric 
 WIMP halo in zoomed in Galactic sized realizations \cite{Abazajian:2020tww,Woudenberg_2024,Hussein:2025xwm,Muru:2025vpz,McKeown:2025zzf}, see also \cite{Hu:2026apn}.
 
Concurrently, neural network-aided simulations have pointed out that the GCE can be a truly diffuse emission as due to DM annihilation, or to a population of very numerous dim point sources \cite{List:2021aer,List:2025qbx}. 
A definitive confirmation on the existence of a new MSP population within the Galactic bulge is expected to come from the detection of these sources through deep radio observations, see e.g.~\cite{Calore:2015bsx,Berteaud:2024rgv,Berteaud:2025ttq,Perez:2026jim}.

If the GCE comes indeed from the cumulative emission of a novel population of MSPs, a  number of dim point-sources, which exact number depends on the assumed luminosity function \cite{Dinsmore:2021nip}, is expected to emerge  with fluxes close to the detection threshold of \fermi-LAT catalogs. 
Statistical methods such as the photon count statistics \cite{2011ApJ...738..181M,Zechlin:1,Mishra-Sharma:2016gis} have been developed to measure point sources below the flux threshold in the catalogs. 
This permits a  quantitative \textit{decomponsition} of all known components of the gamma-ray sky, and to measure the source count distribution  down to about  $\sim 10^{-11}$ ph/cm$^{2}$/s, depending on the energy range. In a given a region of interest (ROI), these techniques assign photons also to unresolved sources, which otherwise could belong to other templates or leave structured residuals. 
However,  when applied to the inner Galaxy, photon count statistics techniques are biased by the mismodeling of Galactic diffuse emission \cite{Chang:2019ars,Leane:2020pfc,Leane:2020nmi,Calore:2021jvg,Manconi:2024tgh}. 
Adaptive template-fitting techniques such as \sky~\cite{Storm:2017arh,Bartels:2017vsx,Song:2024iup} have been demonstrated to mitigate this problems
 by absorbing spectral and spatial mismodeling of the templates used to fit \textit{Fermi}-LAT data, reducing the level of the residual photons down to $2\sigma$ level. 
In \cite{Calore:2021jvg}, a subset of us have already applied adaptive template fitting and photon-count statistics techniques to reconstruct the flux distribution of gamma-ray emitters well below the {\em Fermi}-LAT flux threshold, enforcing a potential sub-threshold point-source contribution to the GCE. 
A subsequent analysis, extended to energies above 10 GeV \cite{Manconi:2024tgh},  has remarkably found that our fits prefer a bulge morphological model over the DM one at high significance, and show no evidence for an additional DM template. 
In \cite{Calore:2021jvg,Manconi:2024tgh} we have established that the combination of these two techniques is necessary to stabilize the photon-count statistics results, and specifically the extracted source-count distribution, in the inner Galaxy. 

In this paper, we re-evaluate the GC as a powerful target for indirect DM searches via gamma rays. 
With respect to earlier constraints \cite{Hooper:2011ti,Ackermann_2012,Fermi-LAT:2017opo,Abazajian:2020tww} we introduce a number of innovations in the GCE analysis.  Specifically, building on the framework we developed in \cite{Calore:2021jvg,Manconi:2024tgh}, we introduce two main novelties.
First, the adaptive template fits are performed with the \sky~code~\cite{Storm:2017arh}, enabling us to fix the spectral shape 
of the DM signal for each  DM mass and final state, while fitting only its normalization, namely the annihilation cross section \sigmav.  In this way we test the evidence for a
combination of both DM and a stellar bulge, for specific DM models. 
Second, we feed the photon-count statistics analysis of the the 1-point probability distribution function method (\onep)~\cite{Zechlin:1} with optimized diffuse emission templates and {\em with} a population of unresolved point sources. 
Point sources with fluxes below the detection may still contribute one or more photons to the observations, thereby adding to the total photon count in the ROI. Their modeling is thus crucial, as their contribution can be comparable in magnitude to other components, such as a potential DM signal, as explored for high latitudes in~\cite{Zechlin:2017uzo}.
The measurement of the source-count distribution, which was a genuine result in \cite{Calore:2021jvg,Manconi:2024tgh}, is a byproduct also of this new analysis, however does not represent a novel result here.
We support our results with a large number of tests on simulated data.

The paper is structured as follows. In Section~\ref{sec:GCE-models} we detail the modeling for the DM and stellar bulge. The methods used to analyze \fermi-LAT data and derive DM constraints are illustrated in Section~\ref{sec:methods}. Our main results are presented in Section~\ref{sec:results}, and supported by discussion and conclusions in Section~\ref{sec:conclusions}. The Appendix~\ref{app:simulations} summarizes the simulations performed to validate our framework.

\section{Dark matter and Stellar bulge modeling}
\label{sec:GCE-models}
The two main hypotheses currently considered to explain the GCE are, on one side, photons produced by particle DM annihilations in the Galactic halo (referred to as \textit{DM} in what follows) and, on the other side, the emission from a new population of MSP-like objects in the Galactic bulge of the Milky Way (\textit{stellar bulge} in what follows). 

Although the bulk of the GCE emission has been demonstrated to be  better explained by a stellar-like bulge emission owing to the observed morphology and energy spectrum (see the most updated analyses in \cite{Song:2024iup,Manconi:2024tgh}), this does not exclude a subdominant contribution from DM.
We  explore the possibility that DM can co-exist with the stellar bulge, by modeling the gamma-ray emission towards the GC as a combination of both contributions.
For this, we build a spatial and spectral template for both components.

\subsection{Dark matter}
In order to compute the expected emission from DM annihilations at the center of the Milky Way, we need to specify a particle DM model, an energy spectrum for the emitted photons, and a density profile.
Under the hypothesis that DM is made of new particles arising at the GeV mass scale as thermal relics of the early Universe, e.g., 
WIMPs, and that they annihilate within the Milky Way halo, photons with GeV energies can be produced as direct annihilation lines, and in the form of a prompt, 
continuous gamma-ray spectrum over several decades in energy by secondary
processes such as the decay of $\pi^0$-mesons, and bremsstrahlung.

The differential gamma-ray flux from DM annihilations per unit energy interval $(E,E+\mathrm{d}E)$ and solid angle $\mathrm{d}\Omega$ in a given celestial direction can be generally computed, for self-conjugated particles, as:
\begin{equation}\label{eq:dmflux}
  \frac{\mathrm{d}\phi_\mathrm{DM}}{\mathrm{d}E \mathrm{d}\Omega} =
  \frac{1}{4\pi} \frac{\langle \sigma v \rangle}{2}
  r_\odot \frac{\rho_\odot^2}{m^2_\mathrm{DM}} \sum_f
  \left(\frac{\mathrm{d}N_f}{\mathrm{d}E} B_f \right) \mathcal{J}(\psi)\,,
\end{equation}
where  $\langle \sigma v \rangle$ is the
thermally averaged self-annihilation cross section times the relative
velocity, $m_\mathrm{DM}$ indicates the DM particle mass, and the position of the solar system in a galactocentric reference frame is $r_\odot$, with the corresponding DM density $\rho_\odot$. The energy spectrum of the emitted photons is ruled by the term  $\mathrm{d}N_f/\mathrm{d}E$, which corresponds to the differential gamma-ray spectrum
yielded by DM annihilation into the standard model final state $f$
with branching ratio $B_f$. Finally, the so-called $\mathcal{J}$-factor is a dimensionless quantity and encodes the dependence on the (squared) DM density profile $\rho(r)$ as a function of the
galactocentric radius $r$:
\begin{equation}\label{eq:j_factor}
  \mathcal{J}(\psi) = \frac{1}{r_\odot} \int_\mathrm{l.o.s.}
  \left( \frac{\rho [r(l)]}{\rho_\odot} \right)^2 \mathrm{d}l(\psi)\,, 
\end{equation}
$l$ being the line-of-sight (l.o.s.) as measured from $r_\odot$. 

We fix the DM model to be a generic WIMP with mass $m_\mathrm{DM}$ in the range from $10$~GeV to $1$~TeV, testing the following specific mass values: $10, 40, 60, 80, 100, 200, 500, 1000$~GeV. 
While in principle other models can be tested within our framework, this choice is driven by the observation that such WIMPs would produce a signal with its peak emission in the energy range in which the GCE is brightest.
For similar reasons, we test two main annihilation channels with $B_f=1$, one hadronic ($b \bar{b}$) and one leptonic ($\tau^{+} \tau^{-}$). Indeed, they both provide, for masses around tens of GeV, an energy spectrum peaked at few GeV, a distinctive feature of the GCE spectrum, and represent thus the most degenerate, and challenging with respect the MSP interpretation, DM models to be tested and constrained~\cite{Calore:2014nla}. 
However, we note that while a number of specific WIMP models have been attempted to be fitted to the GCE spectrum, including combinations of different annihilation channels, they fall short at contextually explaining both the peak and the high energy tail, observed extending up to few tens of GeV, see e.g. the recent studies in  \cite{Hu:2025thq,Koechler:2025ryv,Hooper:2025fda,Roy:2025zvo}.

The spatial density profile is a crucial ingredient to predict the DM signal, and specifically the expected spatial morphology of the emission in the sky. The benchmark models used to investigate the GCE usually assume a DM density $\rho(r)$ parametrized with respect to the radial distance $r$ from the center of the Galaxy. 
In particular, to accommodate the observed photon distribution in the inner few degrees, a generalized Navarro, Frenk and White (NFW) profile \cite{1997ApJ...490..493N} with $\gamma=1.26$ is found to best describe the data in the case of a DM interpretation  \cite{Song:2024iup}. 
However, the DM density of the Milky Way today is only known with sizable uncertainties, which can reach an order of magnitude in the inner few kpc. This level of uncertainty is quantified  through N-body simulations \cite{Schaller:2015mua, Calore:2015oya,2022MNRAS.513...55M,Hussein:2025xwm} and confirmed with empirical approaches based on various astrophysical observables \cite{Benito:2019ngh,Benito:2020lgu}. 
 
We here consider four benchmark profiles derived using updated Milky Way rotation curve data within the framework presented in \cite{Benito:2019ngh} and updated in \cite{Benito:2020lgu}: a NFW profile with $\gamma=1$ (NFW100), a generalized, contracted NFW profile with $\gamma=1.26$ (NFW126), an Einasto \cite{1965TrAlm...5...87E} and a cored, Burkert profile \cite{Burkert_1995}, largely covering the current uncertainties on the shape of the DM halo. 
Among the results of the parameter scan presented in \cite{Benito:2020lgu}, which explored in particular the degeneracy among the local DM density and the scale radius of each profile, we decided to keep the scale radius at $20$~kpc for all of the profiles and to select admissible parameters following the marginal posterior distributions derived by them. 
We fix the solar system distance from the center of the Galaxy to be 8.187~kpc, which makes $\rho_{\odot}$ the only variable quantity for each profile. 
Specifically, following the parametric form of each profile defined in \cite{Benito:2020lgu}, this translates to a local DM density of $0.5\;\mathrm{GeV}/\mathrm{cm}^3$ for all selected profiles. 

Using these parametrizations, the uncertainties in the derived $\mathcal{J}$-factor within a few degrees from the center of the Galaxy are found to be consistent with the ones independently derived using N-body simulations of Milky-Way analogues, see e.g. \cite{Hussein:2025xwm}.

\subsection{Stellar bulge}
If the GCE originates from the cumulative emission of a new population of MSP-like objects,  the spatial distribution of such stellar population and their energy spectrum in gamma rays is needed in order to characterize the expected signal.  

As for the  morphology, similarly to \cite{Calore:2021jvg,Manconi:2024tgh} the spatial template for the stellar bulge component is built from the sum of a boxy-bulge and a nuclear bulge. However, compared to our previous \sky-\onep~ combined analyses~\cite{Calore:2021jvg, Manconi:2024tgh}, we update the boxy-bulge template to follow the so-called Coleman bulge \cite{Coleman:2019kax}, which has been found to be the preferred morphology when the GCE is fully explained by a stellar bulge \cite{Song:2024iup}.   

When analyzing both real and simulated \textit{Fermi}-LAT data, the input energy spectrum is taken to be a power law with an index of $-2.5$, to fully accommodate the high-energy tail of the GCE. However, we note that within the \sky~ fit we leave full freedom for spectral, bin-by-bin modulation, as detailed in the next section. The systematics connected to the choice of the input spectrum for energies higher than 10~GeV have been studied in \cite{Manconi:2024tgh}. 
Instead, for the \sky~ simulation tests described in Appendix~\ref{app:simulations} the spectrum of the stellar bulge is  a power law with exponential cutoff.   
In this case, we adopt the values of the stacked MSP spectrum reported in \cite{McCann:2014dea}, that is, a spectral index of $-1.46$ and a cutoff at 3.6 GeV.
This allows us to test the performance of our method when DM and stellar bulge components exhibit rather degenerate spectral profiles and when our gamma-ray fits are rather agnostic of the spectrum of the stellar bulge.  

The number of the MSPs making up the GCE emission can be derived assuming that the total observed flux comes from this  population, under the hypothesis of a specific luminosity function. 
As demonstrated in \cite{Dinsmore:2021nip}, the number of MSPs explaining the GCE can vary from a few hundred to more than $10^4$, depending on the assumed luminosity function. 
As detailed in what follows,  we leave the normalization of the stellar bulge component free to adjust to the GCE emission, avoiding any correlation to  luminosity models. Therefore, we do not derive any conclusion on the number of MSPs contributing to the GCE.

\section{Methods}
\label{sec:methods}

The \sky~and the \onep~methods are established and well-tested tools to analyze and interpret \textit{Fermi}-LAT data. They have been detailed in multiple previous works,  also when applied together to investigate DM signals at high latitudes, or the inner Galaxy \cite{Zechlin:2017uzo,Calore:2021jvg,Manconi:2024tgh}. 
We present below a summary of both methodologies, recalling the ingredients that enter each fit and stressing the complementarity that they offer in inferring the properties of gamma-ray emission in the inner Galaxy. 
In short, we first fit the \fermi-LAT data in a wide energy range and region of interest using the \sky~method to optimize the model of the inner Galaxy, and specifically to reduce background mismodeling. The optimized templates for all the sky components (diffuse backgrounds, DM and stellar bulge) are then implemented in the \onep~to interpret the \fermi-LAT data within a reduced energy interval and spatial region, and most importantly, including both bright and unresolved sources. 
It is from this second fit that DM constraints are derived. 

With respect to \cite{Calore:2021jvg,Manconi:2024tgh}, computing constraints on DM \sigmav\ requires  the inclusion of both DM and stellar bulge templates within the \onep\ fit to the gamma-ray sky. Extensive simulation tests, which are detailed in the Appendix, support the robustness of these results. 
The statistical framework used to extract the DM constraints is defined in Sec.~\ref{sec:methods:constr}.

\subsection{skyFACT}
\label{sec:SF_intro}

\sky~is a regularized, high-dimensional adaptive template–fitting framework introduced in Ref.~\cite{Storm:2017arh} and subsequently employed in the inner Galaxy to characterize the GCE in Ref.~\cite{Bartels:2017vsx,Song:2024iup}. The property of being adaptive is achieved by augmenting the initial three-dimensional fit model components (one energy and two spatial dimensions) with nuisance parameters per energy bin and spatial pixel while penalizing departures from these prior expectations, thereby absorbing spectral and spatial mismodeling of the templates used to fit \textit{Fermi}-LAT data.

Following the path set by previous works combining \sky~and the \onep~technique in the context of GCE studies~\cite{Calore:2021jvg, Manconi:2024tgh}, our \sky~fit model and statistical setup follow Ref.~\cite{Bartels:2017vsx}. Each gamma-ray emission component is specified by separable inputs for spectrum and spatial morphology. The adopted fit model can be categorized into four broader classes of emission components:
\begin{itemize}
    \item \textit{Galactic diffuse emission}: This category of components is comprised of, \emph{(i)} inverse Compton radiation with spectrum and maps computed for a representative cosmic–ray source and transport scenario using \texttt{DRAGON}~\cite{2011ascl.soft06011M}; and, \emph{(ii)} neutral–pion decay, implemented as three independent gas templates to increase flexibility. The gas templates correspond to rings at 0--3.5, 3.5--6.5, and 6.5--19 kpc, each given by the sum of atomic and molecular hydrogen from the \texttt{GALPROP} public release\footnote{\url{https://galprop.stanford.edu/}}~\cite{Porter:2021tlr}. The gas spectra are fixed to the pion–decay shape from~\cite{2012ApJ...750....3A} and are identical across rings. Lastly, we add the Fermi Bubbles modeled with a uniform brightness template and input spectrum from~\cite{Fermi-LAT:2014sfa}. 
    \item \textit{Point-like and extended discrete sources}: We include all detected, discrete gamma-ray sources listed in the \textit{Fermi}-LAT collaboration's (8-year) 4FGL catalog \cite{Fermi-LAT:2019yla},  adopting the same configuration as in previous works \cite{Calore:2021jvg, Manconi:2024tgh}.
    \item \textit{Diffuse gamma-ray background (DGRB)}: An isotropic term with the best–fit DGRB spectrum of Ref.~\cite{Fermi-LAT:2014ryh} is considered in combination with a spatially uniform template. 
    \item \textit{GCE}: The \sky~fit model is completed with a selection of GCE components as described in Sec.~\ref{sec:GCE-models} depending on the purpose of the respective run. The DM contribution's spatial and spectral input is determined by the selected DM density profile, mass and annihilation channel. In contrast, the stellar bulge component follows the described spatial morphology, while the input spectrum for the fit is a simple power law with index $-2.5$.
\end{itemize}

The dependence of the final optimized diffuse template on the initial set of templates from which the modulation starts was investigated in previous work \citep{Song:2024iup}, see also Appendix~\ref{app:skyfact-sim-results} for an extended discussion.

The \sky~framework is frequentist: fits yield maximum–likelihood values for specified fit model compositions. A full Bayesian model comparison is therefore out of scope, but nested–model tests based on a test statistic are well defined. To this end, we adopt the test statistic and significance definitions employed in previous \sky~studies, cf.~equations in~\cite{Eckner:2024vep}.
\sky's adaptive template fitting approach is based on the minimization of a log-likelihood function. This target function is the sum of a Poisson likelihood and a penalizing likelihood function where the latter exploits regularizing terms whose hyperparameters limit overfitting and control the modulation strength of all nuisance parameters. The optimization process uses the L–BFGS-B algorithm. Our baseline hyperparameters follow \texttt{run5} of Ref.~\cite{Storm:2017arh} for all components except the GCE. The spectrum of both stellar bulge components may float independently in each energy bin, while their morphologies are fixed to the chosen input (no spatial modulation). The DM component is only allowed to be re-adjusted via a global renormalization. 
This means that no spectral and spatial modulation are applied on the DM
template, since we want to test specific DM spectral models to derive consistent bounds.
The global normalization of the DM
template in all the energy range physically corresponds to a rescaling of the cross section. This
is to keep unchanged the energy spectrum predicted by the specific DM  model tested,
e.g. corresponding to the selected DM mass and annihilation channel.

Our main goal is to reduce the mismodeling in the Galactic diffuse emission for the subsequent \onep~fit. 
Indeed, traditional template-based fits with recent Galactic diffuse emission models of the GC region still exhibit residuals of up to 20\% in the 1 -- 10 GeV range \cite{DiMauro:2021raz, Cholis:2021rpp, Pohl:2022nnd}, which can bias the photon-count statistics results~\cite{Calore:2021jvg,Chang:2019ars,Leane:2020pfc,Leane:2020nmi}. 
After the \sky~fit, we extract the best-fit modulated templates per component, and integrate them in the 
energy range between 1.6 and 5.9 GeV to obtain only one energy-integrated template per component. 
We regroup, by adding them together, all Galactic diffuse emission components and extended sources.
This unique template is what we call hereafter ``optimized \sky~diffuse template'', which together with the
optimized DM and bulge templates will serve as a basis for the \onep~fit.
The main novelty of this work derives from the treatment of the GCE components: When fitting the LAT dataset, we always include the DM and stellar bulge components \emph{at the same time} so that the optimized diffuse background templates are bespoke regarding the assumed DM density profile, mass and annihilation channel. Hence, each combination of DM profile, mass and annihilation channel is subject to a dedicated \sky~run.

\subsection{1pPDF} \label{sec:methods:1p}
A crucial component of gamma rays observed by \fermi-LAT originates in point-like and extended discrete sources, which contribute to the total observed photon counts up to tens of percent, depending on energy. 
The number count of detected sources, e.g. in the 4FGL catalog, is found to decrease for fluxes $< 10^{-11}$~ph cm$^{-2}$s, where the
catalog's detection efficiency drops below one \cite{Fermi-LAT:2015otn,DiMauro:2017ing}. Discrete sources with fluxes lower than the detection threshold are too faint to be detected individually, but will still contribute  with one or more photons during the observing time, giving a  contribution to the photon count in the considered region of the sky, that can be of the same order of magnitude as other components, such as e.g. a putative DM contribution~\cite{Zechlin:2017uzo}. 

Photon-count statistics methods \cite{2011ApJ...738..181M,Zechlin:1,Mishra-Sharma:2016gis} have been developed in the last decade in order to measure the collective source-count distribution \dnds~of bright and faint sources, i.e. the distribution of the number of sources $N$ as a function of their observed gamma-ray flux $S$,  integrated in a given energy range $[E_{\rm min}, E_{\rm max}]$. 
These methods exploit the fact that different classes of gamma-ray sources (such as isotropic, diffuse, or discrete sources) are expected to contribute with different statistics to the photon counts. 
In the photon budget of a ROI, this permits to go beyond the catalog flux threshold and assign
photons also to unresolved sources, which otherwise could be assigned to other templates or
leave structured residuals. Specifically, discrete point sources are expected to give a contribution following non-poissonian statistics \cite{2011ApJ...738..181M}.
Applications of these methods range from the measurement of the high-latitude $\mathrm{d}N/\mathrm{d}S$ for extragalactic sources \cite{Zechlin:1,Zechlin:2,DiMauro:2017ing,Lisanti:2016jub} to the inspection of the inner Galaxy emission to unveil the origin of the GCE, see e.g.~\cite{Lee:2015fea,Calore:2021jvg,Manconi:2024tgh}.

We here utilize the \onep~analysis of \fermi-LAT data, as first introduced in \cite{Zechlin:1} and further developed in \cite{Zechlin:2017uzo,Calore:2021jvg,Manconi:2024tgh}. 
At the core of this method is the extraction of the contribution of point sources, and other isotropic and diffuse components based on
the statistical analysis of the probability distribution $p_k^{(p)}$ of the photon counts $k^{(p)}$ in each pixel $p$ of a pixelized map; see ~\cite{Zechlin:1} for the details on the mathematical background.
In short, the \onep~extracts the average source-count distribution \dnds~from the number of $k$-photon sources $x^{(p)}_k$ in each pixel within a given region of the sky, and is sensitive to photon fluxes about one order of magnitude lower than the sensitivity threshold of source catalogs using the same \fermi-LAT dataset. 

When dealing with the inner Galaxy, it has been demonstrated that photon-count statistics methods are not robust when significant, small-scale residuals are left in the region of interest from a mis-modeling of other components, such as the Galactic diffuse emission or the Fermi Bubbles. 
This causes the method to reconstruct a biased $\mathrm{d}N/\mathrm{d}S$, which can show unphysical peaks around the detection threshold of catalogs \cite{Calore:2021jvg,Manconi:2024tgh}, thus reconstructing small-scale residuals as point source populations, and in general, biasing the full statistical fit to the gamma-ray sky \cite{Chang:2019ars,Leane:2020pfc,Leane:2020nmi}. 
Feeding the photon-count statistic fit with optimized background templates obtained within the \sky~framework  has been demonstrated to stabilize the photon-count statistics analysis, and to return robust results for the $\mathrm{d}N/\mathrm{d}S$ of the inner Galaxy both at few GeV, where the peak of the GCE shows up \cite{Calore:2021jvg} and  when studying the GCE high-energy tail at energies larger than 10~GeV \cite{,Manconi:2024tgh}.

In this work, the photon counts are fitted assuming the following model components: a population of isotropic point sources, as quantified by their \dnds; a diffuse, isotropic background emission; a Galactic diffuse emission template; a smooth template  following the stellar bulge; and a smooth template accounting for possible DM emission. 
The Galactic diffuse emission, the stellar bulge and the DM components are taken as the best fits obtained with \sky, integrating the templates within the energy bin 1.6 -- 5.9~GeV and allowing for a global rescaling $A_{\rm gal/stellar/DM}$.  

The \dnds~is modeled in a parametric way covering both the bright and faint source regimes, using a multiple broken power law with a free normalization and two free breaks, leaving also free the indices below and above each break, following the same parameter prior ranges used in~\cite{Calore:2021jvg}.

The number of counts in each map pixel  coming from the diffuse templates $x_\mathrm{diff}^{(p)}$ entering the \onep~fit can be thus written as: 
\begin{equation}\label{eq:xdiff}
  x_\mathrm{diff}^{(p)} = A_\mathrm{gal} x_\mathrm{gal}^{(p)} +
  A_\mathrm{\rm stellar} x_\mathrm{stellar}^{(p)} + 
  A_\mathrm{\rm DM} x_\mathrm{DM}^{(p)} +\frac{x_\mathrm{iso}^{(p)}}{F_\mathrm{iso}} F_\mathrm{iso}\,, 
\end{equation}
where $F_\mathrm{iso}$ indicates the integral flux of the isotropic diffuse emission, which is used  as a sampling, free parameter in the fit.
The $A_\mathrm{gal}$ and $A_\mathrm{\rm stellar}$ rescalings are sampled with flat priors in the intervals [0.1,1.2] and [0.01, 5] respectively. The prior interval for the $A_\mathrm{\rm DM}$ is instead log-flat and is varied for each DM mass to cover multiple orders of magnitude starting from the \sky~best fit normalization, e.g. from [$10^6$, $10^{14}$] ($m_{\rm DM}=10$~GeV) to [$10^{-2}, 10^5$] ($m_{\rm DM}=1$~TeV) in the benchmark case of NFW126 and $b\bar{b}$ annihilation. 
The novelty here  with respect to what developed in \cite{Calore:2021jvg,Manconi:2024tgh} is that we introduce an additional diffuse template to the model components of the \onep~fit (see Eq.~\eqref{eq:xdiff}), to allow for a GCE being a mixture of diffuse emission correlated with a DM signal and a stellar bulge. Within our parameter scan, this also allows us for the first time to quantify the degeneracies among these two components and the bright and faint point sources measured by the \onep. 
We recall that the GCE, as well as being  due  to potentially multiple diffuse emissions, could also receive a contribution from unresolved point sources belonging to the stellar bulge population.

We correct all components entering the fits for the point spread function (PSF) effect as detailed in~\cite{Zechlin:1}. 
The free parameters ${\bf \Theta}$ are 10 in total, respectively 6 for the $\mathrm{d}N/\mathrm{d}S$ parametrization (normalization, 3 indexes $n_1$, $n_2$, $n_3$, and two free breaks $S_{b1}$ and $S_{b2}$), and 4 for the diffuse templates $F_{\rm iso}$, $A_{\rm gal}$, $A_{\rm stellar}$ and $A_{\rm DM}$. 
The fit to the gamma-ray sky is performed using a pixel-dependent likelihood function $\mathcal{L}({\bf \Theta})$.
We remind here that the \onep~ likelihood is pixel-dependent, and thus includes morphological information provided by templates, which enter with an overall normalization (in the specific energy bin) to the fit, as done in standard template fitting analysis. 

We define a posterior distribution $P({\bf \Theta}) = \mathcal{L}({\bf
  \Theta}) \pi({\bf \Theta}) / \mathcal{Z}$, being $\pi({\bf \Theta})$ the prior and $\mathcal{Z}$ the Bayesian evidence. 
We use \texttt{MultiNest}  \cite{2009MNRAS.398.1601F} to sample $P({\bf \Theta})$, with standard configuration, 600 live points and a  tolerance criterion of 0.2.
One-dimensional profile
likelihood functions are constructed for each parameter using the final posterior sample, in order to get prior-independent frequentist maximum likelihood parameter
estimates.
Consistently with previous \onep~ analysis, since the profile-likelihood uncertainty bands are found to be systematically larger than the Bayesian credible intervals, thus providing more conservative conclusions, we adopt them as our reference results.
Additionally, Bayesian model comparison is performed by means of the Bayes factors, which can be directly computed using the nested sampling global log-evidence  $\ln(\mathcal{Z})$ provided by \texttt{MultiNest}. 

\subsection{Data}
\label{sec:fermi-data}

\begin{figure*}[t]
\includegraphics[width = 0.44\textwidth]{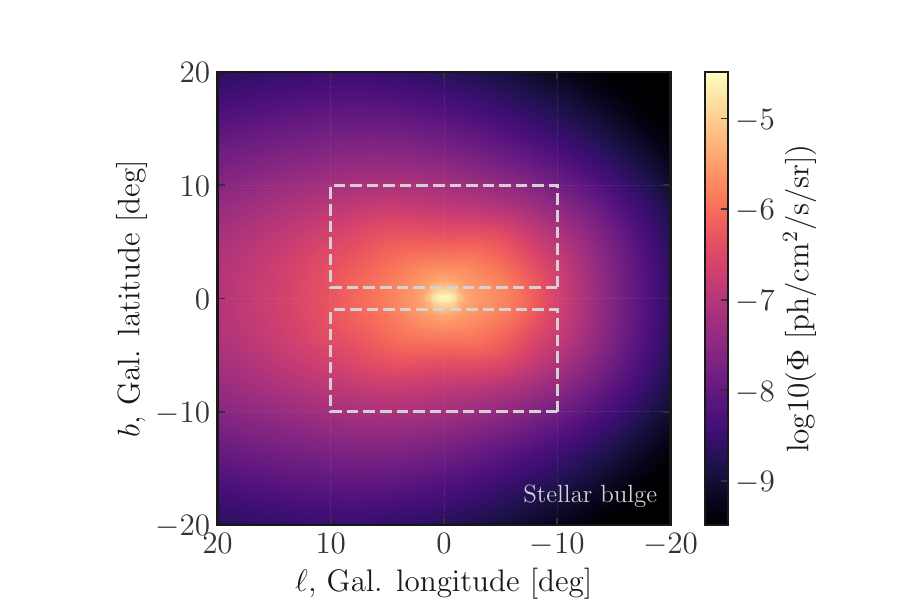}
\includegraphics[width = 0.44\textwidth]{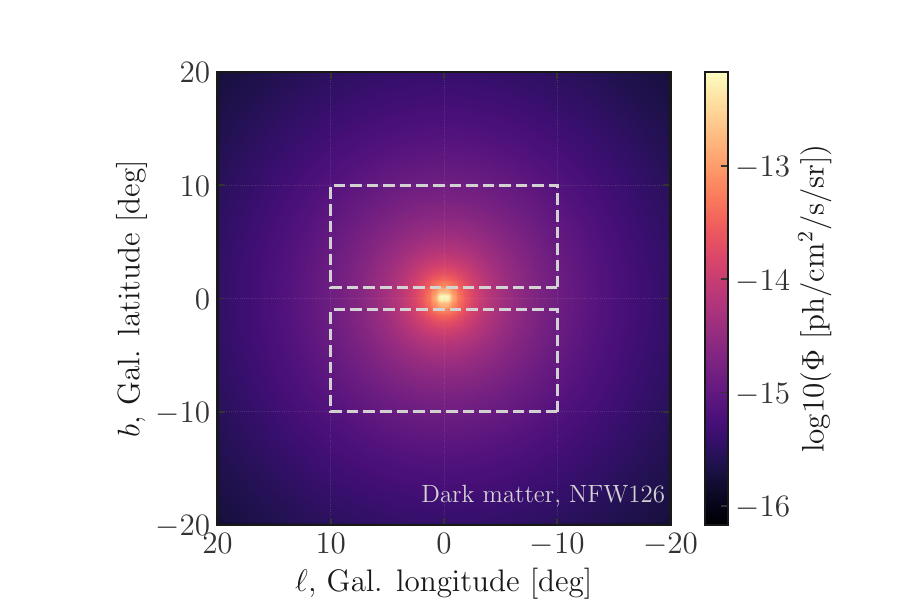}
\caption{Stellar bulge (left) and DM (right) gamma-ray flux maps integrated in the energy bin 1.6 -- 5.9 GeV as obtained with \sky~for the benchmark case of NFW126 density profile and DM of 40 GeV annihilating in the $b\bar{b}$ channel. The flux maps correspond to the best-fit model, and are reported in cartesian coordinates.  The full region of interest (40$\times$40 square degrees) is the one considered for the \sky~analysis, while the dashed lines delimitate the region used to analyze the data with the \onep~method and to derive the DM constraints. The colorbars for the two panels are different, and they both span about five orders of magnitude.}
\label{fig:bulge_dm_maps}
\end{figure*}

For consistency with our previous analyses of the inner Galaxy \cite{Calore:2021jvg,Manconi:2024tgh}, the same dataset of 639 weeks of \textit{Fermi}-LAT data (until 2020-08-27) is used here. 

Regarding the \sky~analysis, this implies utilising \texttt{P8R3 ULTRACLEANVETO} events of \texttt{FRONT+BACK} type from 200 MeV to 500 GeV in 30 logarithmically spaced energy bins and $0.5^{\circ}$ square pixels. We further apply selection cuts only to retain photons with zenith angles $\leq90^{\circ}$ and those that satisfy the standard quality criteria.

As for the \onep,  we concentrate on a single energy bin between 1.6 GeV and 5.9~GeV, and restrict to  \texttt{P8R3 ULTRACLEANVETO} events with the best angular reconstruction (evtype=PSF3), as done in \cite{Calore:2021jvg} for a similar energy interval. Before entering the \onep~ pipeline, events are binned spatially following the HEALPix pixelization scheme \cite{Gorski:2004by} and setting the resolution to $\rm{NSIDE}=128$, corresponding to a mean spacing among bins of 0.4581~degrees. 

The benchmark region of interest (ROI) for the \sky~analysis extends in a square of 40 $\times$ 40 degrees centered at the Galactic center. No latitude mask is applied for the \sky~fit. For the \onep, following the procedure extensively investigated in \cite{Zechlin:2017uzo},  we optimized the ROI to assure statistically stable and robust results, see Appendix~\ref{app:simulations}. The ROI defined as \textit{Inner Galaxy} (a region of 20 $\times$ 20 square degrees again centered at the GC) with a mask of the Galactic plane at $\pm1$ degree has been found to provide statistically sound results, and is used to derive the DM constraints.

The two ROIs adopted in this work are illustrated in the maps in Fig.~\ref{fig:bulge_dm_maps}. The \sky~ROI covers the full  longitude and latitude interval, while the \onep~is identified by the gray dashed lines. 
The models maps illustrate the results for the gamma-ray flux of the stellar bulge and DM components within a benchmark \sky~fit, and will be further commented on in Sec.~\ref{sec:results}.

\subsection{\sky\ and \onep\ combined: dark matter constraints}\label{sec:methods:constr}

Model-independent constraints on thermal relic DM are obtained for two representative hadronic and leptonic channels, $b\bar{b}$ and $\tau^+ \tau^-$ respectively, in terms of the thermally averaged annihilation cross section \sigmav~as a function of the DM mass $m_{\rm DM}$. 
We stress again that, while we explore in this work masses included from 10~GeV to 1~TeV, the developed framework can be extended to explore higher DM masses, for which the DM contribution can still be sizeable with respect to the other components at \fermi-LAT energies.

 To explore possible degeneracies, the DM and stellar bulge components are fitted together, with free normalizations, within both methods, as detailed above.
The \onep, in particular, combines optimized templates obtained with \sky~for the different components of the gamma-ray sky, including DM, stellar bulge and background diffuse emission, and re-fit their free normalizations taking into account also the population of bright and faint point sources within the ROI.
In other words, the \onep~quantifies the role of point sources \textit{simultaneously} to the intensity of the Galactic diffuse emission, of the stellar bulge and, if any, of the diffuse emission due to annihilating DM, since all these components enter the \onep~with an extra, overall free normalization. 
As quantified in \cite{Calore:2021jvg} (see Table I, benchmark setup), 
 13\% of the flux budget of the ROI goes into point sources (bright+faint), 77\% to diffuse backgrounds, and 10\% to diffuse stellar bulge GCE. 
These numbers tell us that the 1pPDF fits to \fermi-LAT data in the inner Galaxy, for the preferred model (which includes a GCE modeled as a stellar bulge) find non-null (and even comparable) emission from both the point source population and the smooth stellar bulge template, in most cases each contributing more than 10\% of the total emission in the analyzed ROI. 
Since sources listed in the 4FGL catalog  account for 7\% of the total  emission, this means that remaining flux assigned to point sources comes from subthreshold ones.  
These quantitative results are possible only when using a method like the 1pPDF, that can measure the source-count distribution of sources below catalog threshold. 
This is relevant for a generic analysis of the \fermi-LAT data, and in particular for our goal of extracting DM constraints. 

The combination of the two methods introduced above brings a significant computational complexity. However,  as we have explicitly shown in \cite{Calore:2021jvg,Manconi:2024tgh}, improving the Galactic diffuse emission model is instrumental for the subsequent \onep~analysis. When running the \onep~ with non-optimized diffuse models, the small-scale residuals left in the ROI could be misinterpreted as point sources, leave more space for additional components and bias the reconstruction of the source-count distribution, as well as the normalization of the other diffuse templates.

The DM limits are extracted from the \onep~analysis of a single, integrated energy bin, chosen to cover an energy interval where the GCE spectrum peaks. As shown in \cite{Zechlin:2017uzo} for high Galactic latitudes, repeating the analysis in different energy bins can improve the sensitivity to lower or higher DM masses. While developing our combined analysis, we explored the potential gain in sensitivity when including information from multiple energy bins at the same time.  
This was achieved by performing a joint fit across different energy bins, each one with its free parameters, except the DM normalization treated as a common parameter. 
Since this approach yielded only a marginal improvement in sensitivity at the cost of a substantial increase in computational time, we chose to adopt the single-energy-bin analysis.

The one-dimensional profile likelihood functions for each parameter, and thus also for the normalization of the DM annihilation cross section, are built from the final posterior \texttt{Multinest} sample, in order to get prior-independent frequentist parameter estimates.  The statistical procedure adopted to build and interpret the construction of the profile likelihood is introduced and described in \cite{Zechlin:1,Zechlin:2017uzo}.
For each parameter of interest, its sampled range is divided into bins, and in each bin we identify the maximum log-likelihood value among all sampled points. The resulting set of maxima as a function of the parameter provides a discrete approximation to the profile likelihood, i.e. an estimate of the likelihood maximized over nuisance parameters at fixed parameter values. The best-fit point is given by the global maximum, and the corresponding error bounds are computed.   If the \texttt{Multinest} run has adequately explored all regions relevant for likelihood extrema, including low-posterior-mass regions that may contribute to the profile likelihood, this approach is expected to provide sufficient sampling accuracy. 
The validity of this approximation was checked explicitly through simulation tests and constant cross-comparisons between the Bayesian \texttt{Multinest} scan and the profile likelihood results. Specifically, this has been verified with different sets of simulations  in \cite{Zechlin:1} (see their Fig. 19 and 20), in \cite{DiMauro:2017ing} and for the DM case at high latitudes in \cite{Zechlin:2017uzo}.

The DM constraints are obtained from the final \onep~posterior sample for each combination of mass and channel separately. 
For each of these combinations, the one-dimensional profile likelihood functions are built from the final posterior sample obtained from the fit to \fermi-LAT data for all the parameters, including the normalization of the DM component $A_{\rm DM}$. 
Since the normalization of the DM component is found to follow a one-dimensional profile likelihood for all the explored cases, the upper limits are derived as  95\% confidence level (CL) limits by using the condition $2 \Delta \ln \mathcal{L}=2.71$. 

An example of such profile likelihood for the \onep~fit to the real sky is illustrated in the Appendix, where we detail  the results of a number of simulation tests. 
Indeed, before proceeding with the analysis of real \fermi-LAT data, we designed a suite of simulations to validate our combined pipeline in order to quantify its robustness against a number of possible systematics. 
The rationale behind each simulation test, performed for both the \sky~and the \onep~methods, is detailed in Appendix~\ref{app:simulations}. 
The results of these simulations demonstrate that: i) in the \sky~fit we are not missing a DM signal at the level of 100\% to 10\% of the GCE; ii) the upper limits obtained with \fermi-LAT data are found to be compatible with null hypothesis tests of the \onep~pipeline within the chosen ROI; iii) the residual mismodeling after the \sky~fit is quantified at the level of $2\sigma$ Poisson noise, and this is found not to affect the robustness of the null hypothesis tests, and thus the final DM limits.

\section{Results}
\label{sec:results}

In this section, we report the main results of our analysis of real \fermi-LAT data, culminating in stringent constraints on the annihilation cross section of thermal DM relics.

\subsection{Optimized diffuse models and dark matter contribution}
\label{sec:results-SF}

\begin{figure}[t]
\centering 
\includegraphics[width =\linewidth]{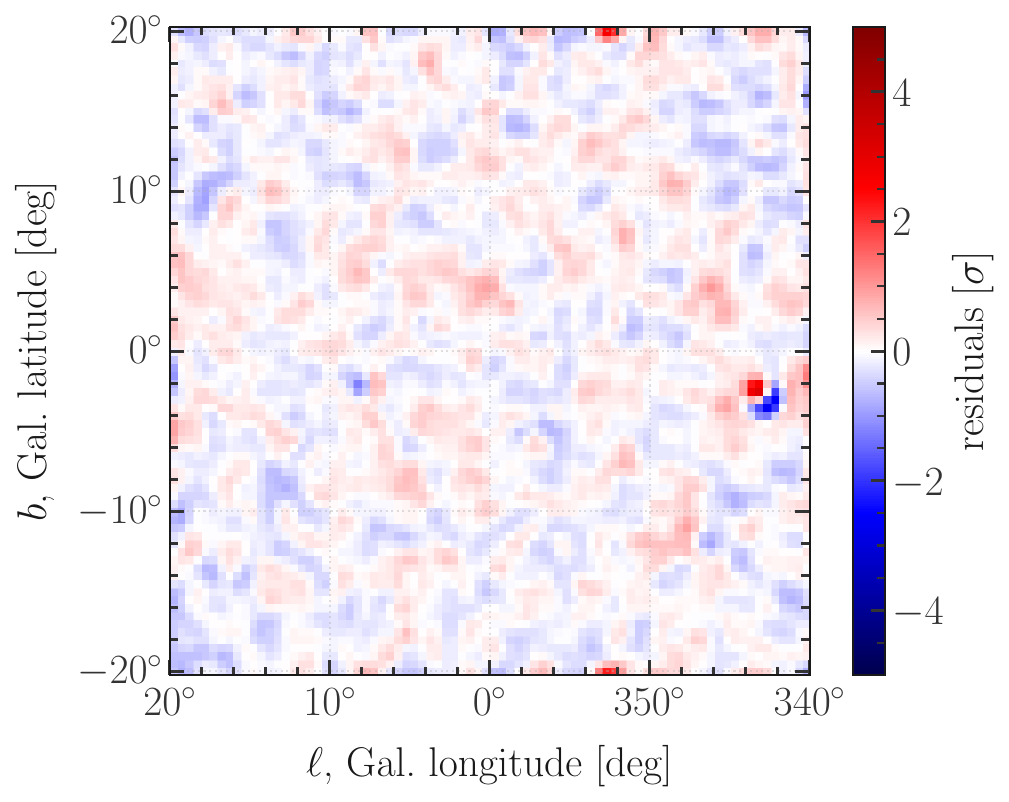}
\caption{\textit{Residuals} of the null hypothesis \sky~fit in the energy range from 1.6 -- 5.9 GeV. The residuals are defined in terms of (data - model)/$\sqrt{\mathrm{data}}$ and smoothed with a Gaussian kernel of size $0.75^{\circ}$.}
\label{fig:skyfact-results-real}
\end{figure}

The starting point for the \sky~fit to the \fermi-LAT data  is the null hypothesis, i.e., a GCE comprised only of the stellar bulge; our \emph{null hypothesis fit model}. The derived optimized templates are the basis for the \onep~simulations to determine an optimal ROI for this analysis part as shown in the Appendix~\ref{app:simulations}.

 To visualize the quality of the \sky~findings,
we display in Fig.~\ref{fig:skyfact-results-real} the fit residuals (regarding the null hypothesis fit model) in our ROI smoothed with a Gaussian kernel of $0.75^{\circ}$. We define the residuals relative to the Poisson uncertainty of the LAT data: (data - model)/$\sqrt{\mathrm{data}}$. There is no visible structure in the residuals that attain values at the $1\sigma$ level. An exception is a localized hot and cold spot at $(\ell, b)\sim(443^{\circ}, -3^{\circ})$ coincident with a very bright extended 4FGL source. However, this  emission is irrelevant regarding the analysis of the GCE with the \onep~method because it is not part of the ROI. 

All subsequent \sky~runs on the LAT data include a DM component on top of the stellar bulge using all combinations of DM density profile, mass and annihilation final states described in Sec.~\ref{sec:GCE-models}. Examples of the optimized GCE components derived during the fit suite are shown in Fig.~\ref{fig:bulge_dm_maps} (NFW126, 40 GeV, $b\bar{b}$). This is a representative example for the \sky~fit outcome to the LAT sky: The stellar bulge is a significantly higher emission than the DM component, which is suppressed by more than eight orders of magnitude. 
We find indeed that very few DM models show evidence above $3\sigma$ of being a necessary addition to a stellar GCE; an example being the tuple (NFW126, 80 GeV, $b\bar{b}$) on which we based our \sky~injection and recovery tests in the Appendix. In fact, we encounter most of these hints for statistical significance around $3\sigma$ when fitting the LAT data with a Burkert DM density profile. The Burkert profile features an extensive core, rendering the DM template almost isotropic, which in turn is a bad fit to the GCE in general and rather degenerate with the DGRB. 

As demonstrated in Appendix~\ref{app:skyfact-sim-results} based on simulated data of our ROI, 
the  normalization of the DM component provides a better indicator for a DM presence in the analyzed data than \sky‘s likelihood values. For $b\bar{b}$ final states, we only find a non-zero normalization for the NFW126 profile for masses of 500 GeV and 1 TeV. They translate to annihilation cross sections of $\langle\sigma v\rangle\sim3\times10^{-27}\;\mathrm{cm}^3/\mathrm{s}$ and $\langle\sigma v\rangle\sim5\times10^{-27}\;\mathrm{cm}^3/\mathrm{s}$, respectively. 
Yet, with the choice of energy range for the \onep~analysis, we are not strongly sensitive to DM emission from such heavy particles. Hence, the obtained best-fit values are significantly below our final upper limits, and do not  conflict with the \onep\ results. 
Thus, these spurious non-zero normalizations are not in conflict with our approach. 
All other combinations of DM parameters lead to annihilation cross sections of $\mathcal{O}(10^{-35}-10^{-34})$ cm$^3$/s, which we essentially consider to be zero. 

As concerns the scenario of $\tau^+\tau^-$ final states, we find a non-zero DM normalization for all tested masses $\geq 60$~GeV when employing the NFW126 profile. The corresponding best-fit annihilation cross sections, however, are at least one order of magnitude lower than the upper limits derived with the \onep~method. 
Therefore, the procedure of setting upper limits in these cases is fully justified. For example, we find the largest cross section for a 1 TeV DM particle, amounting to $\langle\sigma v\rangle\sim1\times10^{-26}\;\mathrm{cm}^3/\mathrm{s}$. The other cross section values are in the same range as the ones quoted for the $b\bar{b}$-channel.
We stress that, as we shown via our simulated data checks on a 100\% stellar bulge GCE, finding a non-zero best-fit normalization is not necessarily equivalent to having found a DM contribution to the GCE. In particular, we found a few realizations of such 100\% stellar bulge GCE mock datasets where a spurious DM contribution was recovered. Yet, the associated integrated flux was always below 10\% of the injected GCE signal. 
In addition, the gamma-ray spectrum induced by $\tau^+\tau^-$ final states is rather hard and peaks around the respective DM mass. Hence, we are potentially probing mainly the high-energy tail of the GCE instead of the 1.6 -- 5.9 GeV energy band. 

Lastly, when we add a DM component to the \sky~fit model, we find an almost negligible impact on the spectral properties of the obtained optimized background components compared to the null hypothesis fit, irrespective of DM mass, annihilation channel or DM density profile. The individual background spectra are compatible with the results reported, e.g., in \cite{Manconi:2024tgh}. The only exception to this statement is the case of 40 GeV DM annihilating into $b\bar{b}$ final states following the NFW126 profile. Here, the template of gas ring III is about 4\% brighter and exhibits a slightly harder power-law index of -2.7 for $E>1$ GeV than all other fits, which seem to prefer an index of -2.75 in that energy range. In fact, gas ring III is the dominant diffuse background component in the 1.6 to 5.9 GeV energy range with an integrated flux of $4.3\times10^{-6}$ ph/cm$^2$/s/sr in the null hypothesis fit (normalized to the angular size of our ROI). As this gas ring is the brightest gamma-ray emission source in our ROI, its slightly enlarged normalization has repercussions on the other components, specifically on the sub-dominant ones like the Fermi Bubbles. Their normalization is lower by a factor of two compared to all of the other fits (that predict an integrated flux of around $3.1\times10^{-7}$ ph/cm$^2$/s/sr in the 1.6 -- 5.9 GeV band). In any case, the normalization of the DM component is zero for this run, so that the observed differences does not impact our limits.

\subsection{Dark matter constraints}
\label{sec:results-1ppdf}

\begin{figure*}[t]
\centering 
\includegraphics[width = 0.48\textwidth]{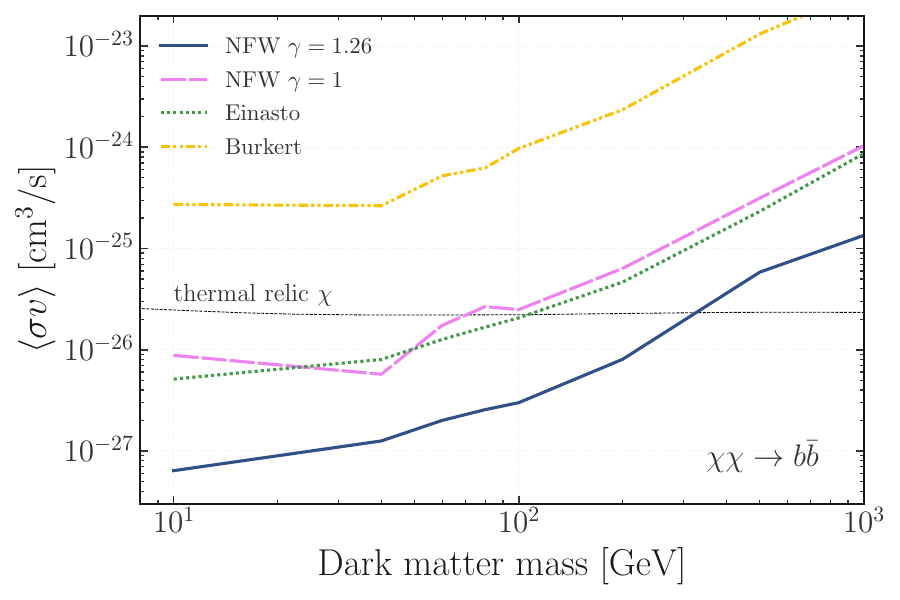}
\includegraphics[width = 0.48\textwidth]{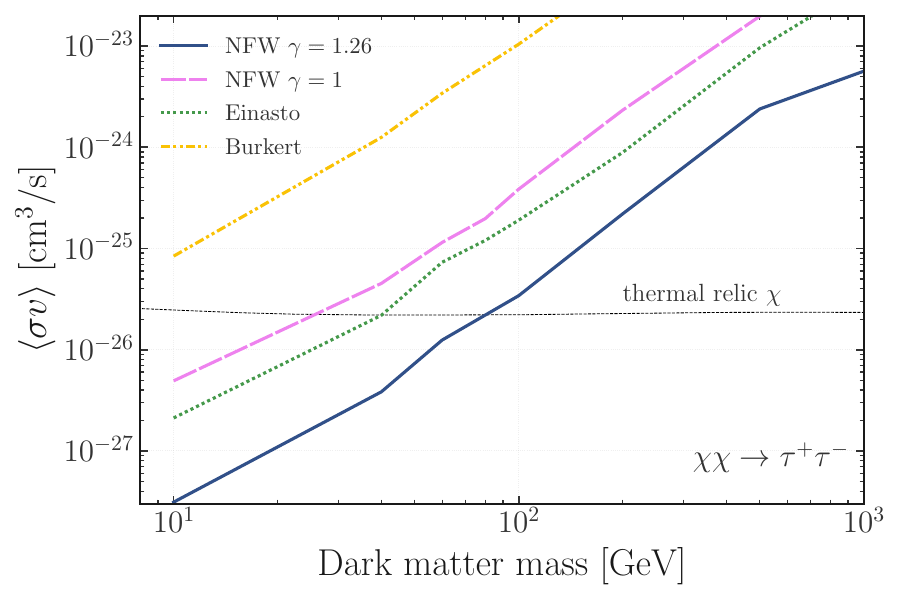}
\caption{\textit{Constraints to thermal relic DM obtained in this work using \fermi-LAT gamma rays from the GC.} 
The upper limits (95\% C.L) on the annihilation cross section $\langle \sigma v \rangle$ as a function of the DM mass are shown
for  the $b \bar{b}$ ($\tau^+ \tau^-$) channel  in the left (right) panel. 
We show the results obtained when varying the Milky Way DM halo profile among the four benchmarks considered: a contracted NFW profile (NFW126, solid blue), a classical NFW (NFW100, dashed magenta), Einasto (dotted green) and the cored Burkert profile (yellow dot dashed). The thermal relic cross section as estimated in \cite{Steigman:2012nb} is also indicated with a thin black line.}
\label{fig:results_sigmav_profile}
\end{figure*}

Armed with the insights on the properties of the gamma-ray sky discussed above, and the extensive simulation tests detailed in the Appendix, we run -- for each annihilation channel and DM mass separately -- the \onep~analysis of the \fermi-LAT data in the 1.6--5.9~GeV energy bin while using the optimized background, stellar bulge and DM templates resulting from the \sky~fit.
We remind that a region 20 $\times$ 20 square degrees centered at the Galactic center, with a Galactic plane mask excluding the inner $\pm 1$ degree, is considered.

By scanning all the channels and DM masses considered, the \onep~analysis consistently finds well-defined, single-sided profile likelihoods for the $A_{\rm DM}$ parameter, from which 95\% CL upper limits are extracted as detailed in Sec.~\ref{sec:methods:constr}.
In other words, despite the fact that the normalization of the DM and bulge templates is allowed to independently vary within the \onep~fit, we do not find a single configuration in which the DM component is detected significantly, consistently with the \sky~results. Conversely, the  parameter scan reveals that the space left for DM in terms of gamma-ray counts is confined to normalizations that are compatible with the null hypothesis tests, see Appendix~\ref{app:simulations}. 
By inspecting the \dnds~extracted in each \onep~fit, we find that the results for the different DM masses and annihilation channels are fully compatible within statistical uncertainties with the \dnds~of the inner Galaxy presented in \cite{Calore:2021jvg} for a similar ROI and energy range.

The two panels of Fig.~\ref{fig:results_sigmav_profile} illustrate the main result of this paper: the constraints on thermal relic DM obtained using the \fermi-LAT observations of gamma rays from the inner Galaxy, presented in terms of the annihilation cross section $\langle \sigma v \rangle$  as a function of the  mass, see Eq.~\eqref{eq:dmflux}. 
The left (right) panel shows the results for the $b \bar{b}$ ($\tau^+ \tau^-$) annihilation channel. 
The four lines correspond to the four DM profiles investigated: a contracted NFW with $\gamma=1.26$ (blue line), a standard NFW (dashed magenta), an Einasto (dotted green) and a cored, Burkert profile (yellow dot-dashed).
As expected, the most constraining upper limits are obtained when assuming a contracted NFW126 profile, which predicts a much larger DM density and thus a higher signal in the ROI. The standard NFW and the Einasto profiles provide similar results across all the DM masses and channels tested, with a maximal difference of about a factor of two for the leptonic channel. 
The upper limits for the cored, Burkert profile are found to be  more than two orders of magnitude higher with respect to the ones derived for the contracted NFW126 profile, and more than one order of magnitude higher when compared to the Einasto and NFW profiles. 
When comparing to the thermal relic cross section as estimated in \cite{Steigman:2012nb}, our limits for the Burkert profile are still one order of magnitude higher along all the mass range. 
On the other hand, in the case of the NFW126 (NFW) profile our results test the thermal relic cross section up to around 300~GeV (100~GeV) for the hadronic channel, and up to 80~GeV (30~GeV) for the leptonic channel.

Despite the fact that we focus on a generic WIMP model, our framework allows us to inspect the consequences of our results for some specific models, for example for Higgsino-like DM, which has been recently the focus of a number of investigations based on \fermi-LAT data, see e.g.~\cite{Dessert:2022evk}. 
The Higgsino is a well-motivated, minimal DM candidate expected with a thermal mass of around 1 TeV, which produces a line-like signature at energies equal to its mass, accompanied by a continuum gamma-ray signal at lower energies \cite{Hisano:2006nn, Cirelli:2007xd}. 
We verified that the prompt spectrum from thermal Higgsino annihilation (prepared adopting the tabulated prompt spectra of \cite{Arina:2023eic}) is consistent, within 10\% in the energy range from 1.6 to 5.9 GeV, with the case of a DM mass of 1 TeV for the $b\bar{b}$ channel within our framework. At a given annihilation cross section, the $b\bar{b}$ gamma-ray spectrum is about a factor of 2.4 larger than the one of the thermal Higgsino. Subsequently, we conducted \sky~fits for all four DM density profiles in combination with the Higgsino gamma-ray spectrum to validate that the obtained optimized diffuse templates are compatible at the $2\sigma$ level with the respective ones derived for the $b\bar{b}$ scenario with 1 TeV DM mass. This comparison was positive for all profiles and, therefore, we can directly use the $b\bar{b}$ upper limit at 1 TeV in the left panel of Fig.~\ref{fig:results_sigmav_profile} to quantify the thermal Higgsino upper limit. It is about a factor of two larger than the $b\bar{b}$ one.  
This is about a factor of 10 less constraining than the upper limits obtained in \cite{Dessert:2022evk} using 14 years of \fermi-LAT data at energies above 10~GeV. Yet, this conclusion does not come as a surprise since our constraints are derived from a narrow energy window where the spectral shape of the DM annihilation signal cannot provide additional constraining power, in contrast to the cited study.

\begin{figure*}[t] 
\centering
\includegraphics[width = 0.48\textwidth]{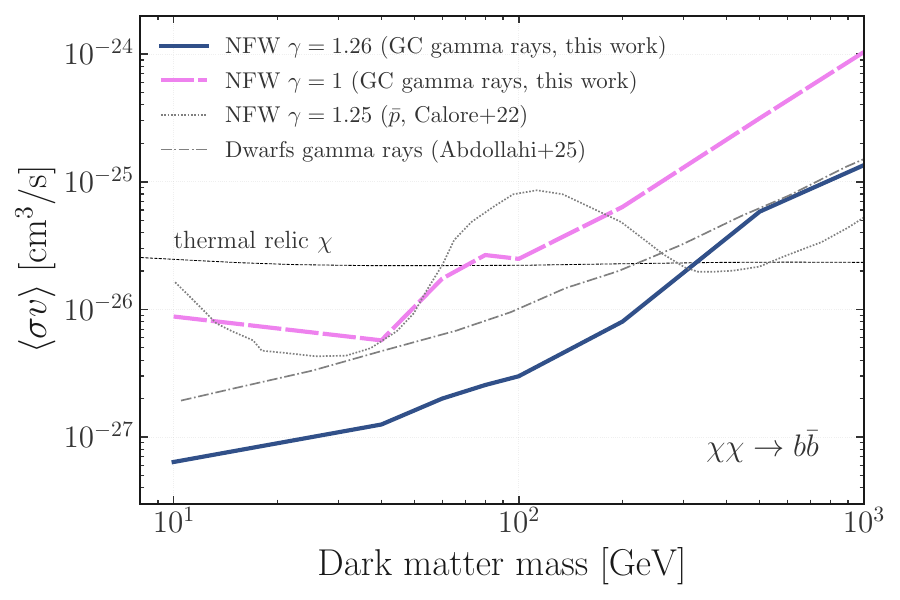}
\includegraphics[width = 0.48\textwidth]{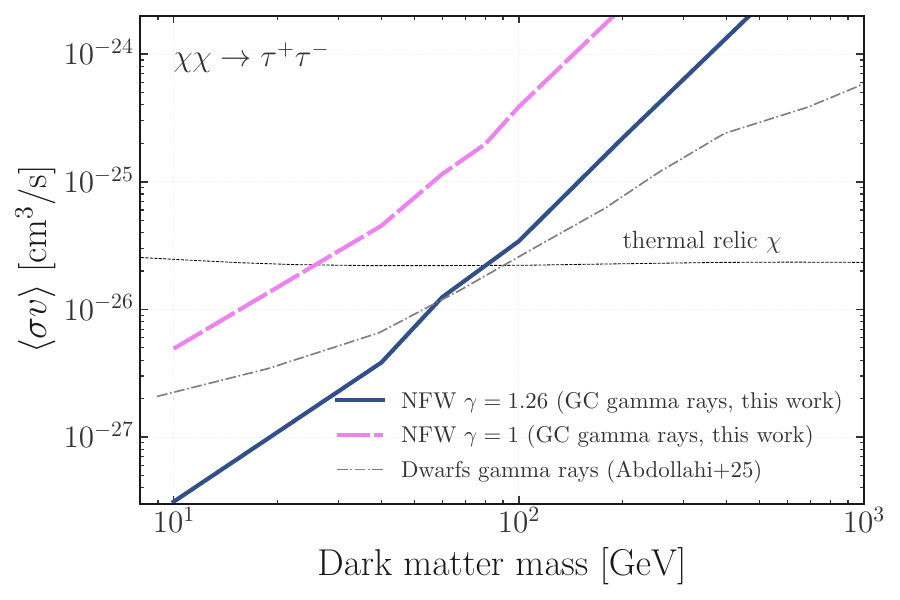}
\caption{\textit{Comparison of the results obtained in this work with complementary multimessenger DM constraints.}  The left (right) panel illustrates the upper limits obtained using the GC gamma rays and assuming contracted NFW profile ($\gamma=1.26$, solid blue) and the  classical NFW profile ($\gamma=1$, dashed magenta) for the  
$b \bar{b}$ ($\tau^+ \tau^-$) channel. These are compared with the upper limits obtained with a similar contracted profile, but using the local cosmic ray $\bar{p}$ fluxes in Calore+22 \cite{Calore:2022stf}, as well as the constraints obtained combining multiple gamma ray observations of dwarf galaxies in Abdollahi+25 \cite{Fermi-LAT:2025fst}.  }
\label{fig:results_sigmav_nfw}
\end{figure*}

\section{Discussion and Conclusions}
\label{sec:conclusions}
In this work, we have scrutinized the space left for signals from weak-scale DM particle 
annihilations in \fermi-LAT data from the inner Galaxy region, and in the presence of a 
stellar-like bulge emission. 
This region has been often dismissed for DM studies because of the evidence for the GCE, 
hindered by uncertainties in the modeling of astrophysical components. 
We here demonstrated that it is nonetheless possible to account for degeneracies between 
the DM signal and the astrophysical modeling, including the stellar bulge. 
For the first time in a fully self-consistent way, we fit the gamma-ray data with a mixed model comprising a DM signal and 
a stellar bulge, both potentially contributing to the excess.
We also account for faint point sources in the region of interest through photon-count statistical 
methods.
Our guiding principle has been to build a consistent description of all known components of the gamma ray sky, notably the faint point sources, and deliver results resilient to the known biases that affect the analysis of gamma rays toward the inner Galaxy. 
The computational complexity brought by the combination of two state-of-the-art methodologies we have applied, \sky~ and the \onep, owes to
the precise description of each known component of the gamma-ray sky, and aims at addressing known bias when modeling them.
The hybrid approach developed in previous papers enabled us to test the presence of DM signals
for different masses, annihilation final states and spatial distributions in the Milky Way halo. 
For all combinations tested (except for very few cases), no evidence for DM is found in the 
\sky~fits of gamma-ray count maps. Through the sub-sequent fit with the \onep, we then derived 
95\% C.L.~upper limits on the annihilation cross section.

The robustness of our results has been supported by a long series of tests, performed on 
simulated \fermi-LAT data and reported in the Appendix.
In particular, we demonstrated through simulations that the mis-modeling induced by the \sky~fit on the optimized diffuse model is at the level of 2$\sigma$, and that this systematic uncertainty does not affect the final upper limits derived with the \onep~analysis.

Our work resulted in stringent limits on particle DM for masses $\lesssim  300$ GeV from gamma-ray data.
The main theoretical systematic uncertainty is related to the DM profile in the Milky Way, which induces
a variation of $\mathcal{O}$(100) in the final limits, regardless of the mass.
Differently from a previous analysis targeting a similar ROI~\cite{Abazajian:2020tww}, our work 
includes sub-threshold point sources in the sky model. 
Moreover, the standard template fit performed in~\cite{Abazajian:2020tww}, does not 
require the DM spectrum to be physical since the fit is done independently 
in each energy bin. This approach hinders the consistency of the results and unavoidably ignore 
degeneracies between the DM model (spectrum and morphology) and the stellar bulge.
We also notice that the milder dependence on the DM profile in their case is due
to the choice of a cored distribution very much different from our Burkert one, and resembling 
the standard NFW beyond a core of 1~kpc.

In Fig.~\ref{fig:results_sigmav_nfw}, we compare our results with a selection of recent, complementary indirect DM constraints.
Traditionally more robust limits from gamma-ray data originate from the null-detection of gamma-ray signals from the direction of dwarf spheroidal galaxies, albeit mis-modeling of the Galactic diffuse emission at the dwarfs' locations can sensibly weaken the limits~\cite{Calore:2018sdx,Alvarez:2020cmw}. We report in Fig.~\ref{fig:results_sigmav_nfw} (gray dot-dashed line), 
the latest combined constraints on DM annihilation from dwarf spheroidal galaxies using data from five gamma-ray observatories (\fermi-LAT, H.E.S.S., MAGIC, VERITAS, and HAWC)~\cite{Fermi-LAT:2025fst}.
For the $b \bar{b}$  channel and up to a DM mass of 500~GeV, the upper limits obtained from the inner Galaxy gamma rays for the NFW126 profile are more sensitive by a factor of about three with respect to these constraints. As for the  harder-spectrum $\tau^+ \tau^-$ channel, our results are competitive only for masses lower than 50~GeV, since the dwarfs' analysis also exploits ground-based gamma-ray detectors, which extend to higher energies.
A very marginal excess relative to background at a 2--3 $\sigma$ (local) level was found in previously combined LAT 
analysis of dwarf spheroidal galaxies~\cite{McDaniel:2023bju}. This excess was not re-confirmed in 
subsequent analyses, and also our limits are compatible with it being due to background fluctuations.

Our constraints are also highly competitive with limits from other messengers and wavelengths: Antiprotons 
offer a clean way to constrain DM signals with high-precision AMS-02 data. 
In Fig.~\ref{fig:results_sigmav_nfw}, we report a comparison with antiproton upper limits obtained by using a similar contracted profile~\cite{Calore:2022stf} (gray dotted line). 
We note that the results of \cite{Calore:2022stf} for the standard NFW profile are very similar to the one reported in Fig.~\ref{fig:results_sigmav_nfw}, as the local antiproton flux is not as sensitive as the inner Galaxy gamma rays to the inner DM profile. 
With respect to the constraints obtained using the contracted NFW126 profile, for masses lower than 200~GeV our results are found to be at least a factor of five more constraining, reaching more than one order of magnitude at 10~GeV and at 100~GeV. 
For masses higher than 300~GeV antiprotons are found to be more constraining. This is expected, as our analysis considers only gamma rays of a few GeV of energy, while the analysis of \cite{Calore:2022stf} exploits $\bar{p}$ measured up to hundreds of GeV.
Moreover, our limits below $\sim$ 100 GeV and for the $\gamma=1.26$ halo profile are comparable with radio bounds on the DM synchrotron emission from the analysis of ASKAP continuum image of the Large Magellanic 
Cloud~\cite{Regis:2021glv}, which, on their side, suffer from uncertainties related to cosmic-ray propagation in the system and magnetic field configuration.

While indirect detection limits are typically agnostic about the specific particle physics model, 
we have also compared them with expectations from high-priority WIMP targets, such as minimal Higgsino DM (but see also the models explored in e.g.  \cite{Baumgart:2025dov,Aghaie:2025iyn,Safdi:2025sfs}).
In this case, we are less constraining than a previously dedicated \fermi-LAT data analysis, explicitly 
targeting this model~\cite{Dessert:2022evk}.
Indeed, we recall that our upper limits are set by a tiny energy band outside  the interesting (and most sizable) part of the Higgsino spectrum.
Analogously, our methodology is general enough to be easily re-adaptable to different DM spectral shapes or masses, 
and can therefore test broad classes of DM particle models.
 
Our work remains confined to testing the ``standard'' spherically symmetric DM profile and their 
possible degeneracies with the stellar bulge. 
However, our methodology can be adapted to include non-spherical DM profile and to re-assess the
evidence of DM in the presence of significant asymmetries in its morphology, as recently suggested~\cite{Muru:2025vpz}.

Finally, it is a fact that one additional limitation of this framework is the narrow energy range of the
analysis. We recall that the choice of the $\sim$ 2 -- 5 GeV energy bin has been dictated by our efforts to provide conservative
results. Nonetheless, extension to higher energies is technically possible~\cite{Manconi:2024tgh}, but it 
does require further simulations of the reconstruction of DM signals at these energies. 
Definitely, the most suitable instrument for constraining DM at masses above 100 GeV will be the Cherenkov Telescope Array Observatory (CTAO), whose sensitivity predictions have demonstrated a great potential for DM searches~\cite{CTA:2020qlo, CherenkovTelescopeArray:2023aqu, CTAConsortium:2023yak, CTAO:2024wvb, CTAO:2025gdd, Baumgart:2025dov}. In the next few years, during its early-science phase, CTAO may already probe the thermal DM landscape for masses beyond 100 GeV \cite{Abe:2025lci}.

\medskip

\begin{acknowledgments}
This work has been done thanks to the facilities offered by the Univ.~Savoie Mont Blanc - CNRS/IN2P3 MUST computing center. 
Silvia Manconi acknowledges the support of the French Agence Nationale de la Recherche (ANR) under the grant ANR-24-CPJ1-0121-01, and European Union's Horizon Europe research and innovation program for support under the Marie Sklodowska-Curie Action HE MSCA PF–2021,  grant agreement No.10106280, project \textit{VerSi}. 
FD thanks the Department of Theoretical Physics of CERN, where part of this work for carried on, the support from the research grant TAsP (Theoretical Astroparticle Physics) funded by Istituto Nazionale di Fisica Nucleare (INFN), and from the Research grants The Dark Universe: A Synergic Mul- timessenger Approach, No. 2017X7X85K, funded by the Miur.
This publication is supported by the European Union's Horizon Europe research and innovation programme under the Marie Skłodowska-Curie Postdoctoral Fellowship Programme, SMASH co-funded under the grant agreement No. 101081355. The operation (SMASH project) is co-funded by the Republic of Slovenia and the European Union from the European Regional Development Fund.
\end{acknowledgments}

\appendix

\section{Simulation tests}\label{app:simulations}
When interpreting the inner Galaxy gamma-ray emission observed by \fermi-LAT within a given methodology, some crucial questions arise: is the diffuse background mismodeling affecting the results, such as the measured DM flux and the derived DM limits? Are the upper limits obtained with real data compatible with the null hypothesis?
In numerous past works \cite{Chang:2019ars,Leane:2020nmi,Leane:2020pfc}, these questions have been addressed through simulation and injection tests, unveiling often limiting systematics that cast doubts on the robustness of the real sky results.

\subsection{Diffuse mismodeling and DM injection and recovery tests}

\subsubsection{\sky~simulation methods}
Despite the ability of adaptive template fitting via \sky~to reduce the fit residuals and redistribute them among the model components, the resulting optimized templates are not entirely free of a certain degree of mismodeling because, \textit{(i)}, the fit model is not guaranteed to contain all physically relevant gamma-ray emission components and, \textit{(ii)}, the choice of the \sky~hyperparameters determines which components may be modulated and at what level \cite{Song:2024iup}. 

Since in this work the \sky~analysis provides the subsequent \onep~runs with an optimized set of gamma-ray emission templates, we investigate via synthetic data what degree of residual mismodeling we can expect -- and ultimately how this mismodeling impacts the final DM upper limits. This will help us to judge the quality of the optimized templates obtained from the LAT dataset and whether we can trust the eventually derived DM constraints. Secondly, analyzing synthetic data clearly tells us the sensitivity of \sky~to each fit model component when engaging in injection and recovery tests. Such tests are necessary to verify that we are theoretically able to detect a DM signal in the LAT dataset that would be otherwise excluded by the final upper limits we will set with the \onep~analysis. In this section, we outline our rationale for simulating the synthetic data to perform these injection and recovery checks.

\paragraph{Preparation of simulated LAT data.} The general idea is to simulate data from the \sky~fit model composition described in Sec.~\ref{sec:SF_intro}. To this end, we prepare three-dimensional flux models from the spectral and spatial input for each respective component and generate photon count maps convolved with the LAT's instrument response functions via the \textit{Fermi Science Tool}'s routine \texttt{gtmodel}. Spatial and spectral binning of the created count maps are the same as for the real \fermi-LAT data (see Sec.~\ref{sec:fermi-data}).

We simulate the \fermi~Bubbles adopting the model of Ref.~\cite{Fermi-LAT:2017opo}, which uses two different morphological and spectral templates above and below $|b|=10^{\circ}$. This choice is different from the unique template implemented in the \sky\ fit model. This difference in the sky simulation and reconstruction toolkits is an effective way to account for an imperfect knowledge of the Bubbles, which is likely the case given that all currently available models for the \fermi~Bubbles are fundamentally data-driven.

To model the GCE in our test suites, we pick a single representative DM configuration defined by a generalized NFW profile with inner slope $\gamma=1.26$, a DM particle mass of 80 GeV, which annihilates into $b\bar{b}$ final states producing gamma rays. The motivation for picking exactly this configuration is a mild $\sim3\sigma$ evidence for such a DM particle that we uncovered in the LAT dataset (more details in Sec.~\ref{sec:results-SF}). The stellar bulge components follow their spatial profiles discussed earlier in Sec.~\ref{sec:GCE-models} with a power law with exponential cutoff spectrum. This choice of spectral profile deliberately puts us in a situation where stellar bulge and DM GCE components are spectrally degenerate. Our simulations, thus, probe the performance of \sky~in a pessimistic scenario.

\paragraph{Definition of test cases.} Our injection and recovery tests will always feature all non-GCE background components, such as the Galactic diffuse emission, DGRB and 4FGL catalog sources. Instead, the different test scenarios are characterized by their GCE composition. To fix a scale for the GCE itself, we resort to the integrated flux from 1.6 to 5.9 GeV of $\phi_{\mathrm{GCE}} = 2.2\times10^{-7}\;\mathrm{cm}^{-2}\,\mathrm{s}^{-1}\,\mathrm{sr}^{-1}$ we obtained from the \sky~null hypothesis fit (only stellar bulge) to the LAT dataset (see also Sec.~\ref{sec:results-SF}). We use the results of this null hypothesis fit to additionally fix the relative difference in flux of the boxy bulge and nuclear stellar cluster in all simulated datasets; the latter is dimmer than the boxy bulge by a factor of 2. 

We consider four different cases to define the GCE composition with respect to $\phi_{\mathrm{GCE}}$ and the fractional contribution of each GCE component in the energy range from 1.6 to 5.9 GeV:
\begin{enumerate}
    \item 100\% stellar bulge, 0\% DM: This scenario corresponds to the null hypothesis of this study. In this setting, we will examine what level of residual diffuse background mismodeling is present in the modulated template.
    \item 90\% stellar bulge, 10\% DM: This case features a weak DM signal in the data, which corresponds to a cross-section of $\langle\sigma v\rangle=5.5\times10^{-27}\;\mathrm{cm}^{3}\,\mathrm{s}^{-1}$ for the selected DM particle parameters. This value is not excluded by complementary indirect search results (c.f.~Fig.~\ref{fig:results_sigmav_nfw}).
    \item 50\% stellar bulge, 50\% DM.  The required annihilation cross section amounts to $\langle\sigma v\rangle=2.8\times10^{-26}\;\mathrm{cm}^{3}\,\mathrm{s}^{-1}$, roughly the thermal value. Such a scenario is within the current reach of indirect DM searches and excluded by the non-observation of gamma rays from the Milky Way's dwarf spheroidal galaxies \cite{Fermi-LAT:2025fst}.
    \item 0\% stellar bulge, 100\% DM: Lastly, we consider that 100\% of the GCE is comprised of emission from DM annihilation, requiring twice the cross-section mentioned in the previous case. 
\end{enumerate}
For all scenarios, we generate $\mathcal{O}(10)$ Poisson realizations of the simulated data and analyze them with \sky. From the results, we will assess the expected significance of the GCE components, as well as their reconstructed fluxes in comparison to what was injected.

\subsubsection{Results}
\label{app:skyfact-sim-results}

\paragraph{Residual background mismodeling.} The numerous nuisance parameters controlled by hyperparameters render \sky~a powerful tool to significantly reduce the presence of background mismodeling, but not perfectly. We explore the residual level of mismodeling by studying the simulated GCE scenario, which contains 100\% stellar bulge emission and no DM. This case corresponds to our null hypothesis, which we tested and verified for the LAT dataset (see Sec.~\ref{sec:results-SF} for details). The goal here is to quantify the level of residual background mismodeling after the \sky~fit in order to correctly propagate this systematic uncertainty in the final upper limits on DM.

\begin{figure}[t]
     \includegraphics[width=\columnwidth]{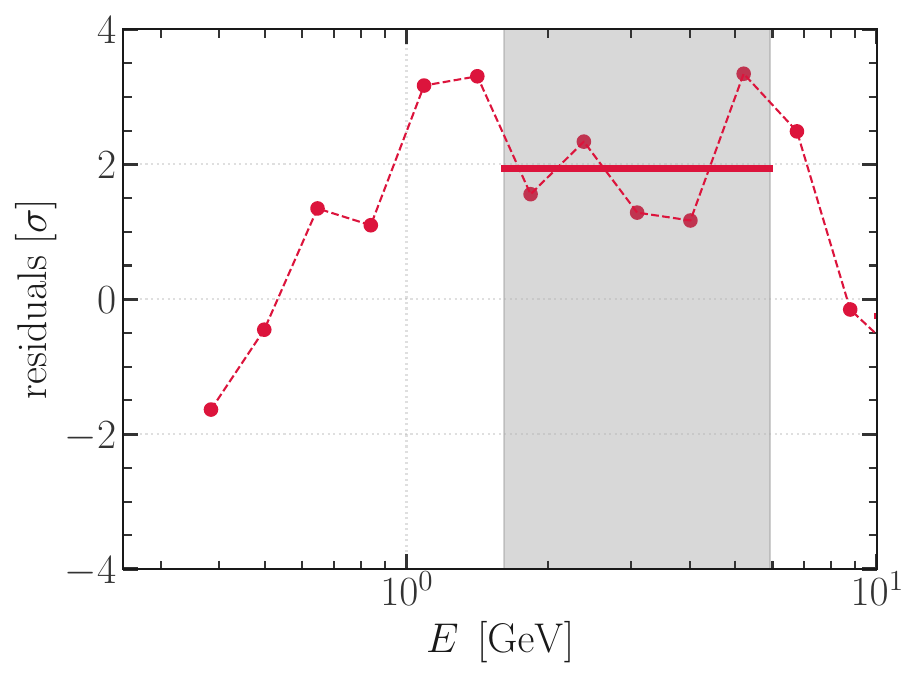}
      \caption{\textit{Performance of \sky~on simulated data: residual background mismodeling.} We illustrate the spectral residuals per energy bin shown as red data points. These residuals were derived for a single realization of the GCE scenario featuring 100\% stellar bulge and 0\% DM emission. The displayed residuals are defined as (data - model)/$\sqrt{\mathrm{data}}$.
      The thick red line marks the residual obtained for the integrated energy range we consider in the \onep~analysis. The gray band denotes the energy range considered in the \onep~ analysis part.
      }
      \label{fig:skyFACT-mismodeling}
\end{figure}

This requires evaluating how well the recovered emission model resembles the injected one, and quantifying the systematic uncertainties associated with the single energy-integrated \sky~template later used in the \onep~fit. The findings are displayed in Fig.~\ref{fig:skyFACT-mismodeling}. This plot shows the obtained spectral residuals of the Galactic diffuse emission, that is, the sum of all components belonging to this class, including the DGRB and reconstructed extended discrete sources from 4FGL, for a representative realization of this GCE composition scenario. The residuals are stated in terms of significance defined as (data - model)/$\sqrt{\mathrm{data}}$, where ``data'' refers to the sum of simulated components and ``model'' denotes the sum of the respective recovered components. This measure is related to the uncertainty of Poisson-distributed data. Across the considered energy bins, the reconstruction can lead to underfitting the synthetic data at the $3\sigma$ level. However, restricted to the energy range between 1.6 and 5.9 GeV,  which we will later pick out to perform the \onep~runs, the integrated residual mismodeling is at the level of $2\sigma$ (horizontal red line and shaded region). 
We  also inspected the obtained spatial residuals, generally finding small-scale residuals and mismodeling at the $2\sigma$ level maximally and sometimes larger residuals localized around the position of the brightest 4FGL catalog sources (c.f.~Fig.~\ref{fig:skyfact-results-real}). 

In summary, while we confirm that the background optimization via \sky~is not perfect, the residual level of background mismodeling in the optimized diffuse template is at the level of $2\sigma$ Poisson fluctuations. Larger residuals may occur related to bright point-like sources. These are, however, measured by the \onep~method in terms of the source-count distribution, which we have verified returns a source population well compatible with the dedicated catalog analysis in the bright source regime, and down to the catalog's flux sensitivity. 

We note that in an earlier \sky-based work \citep{Song:2024iup}, we already investigated the dependence of the optimized diffuse template on the initial seed from which the modulation starts. There, \sky~was run on the same inner-Galaxy region using three qualitatively different background template sets (GALPROP-based from \cite{Cholis:2021rpp}, ring-based from \cite{Pohl:2022nnd}, and the original skyFACT \texttt{run5} composition we use throughout this work) with the same modulation approach. The resulting ``optimized'' diffuse models are not identical because the solution depends on the adopted regularization and hyperparameters, and a seed that is too discrepant cannot necessarily be repaired within the allowed modulation ranges. However, as a heuristic cross-check, we checked that the \sky-optimized diffuse templates obtained from these different seeds are mutually compatible at the $\mathcal{O}(<2\sigma)$ level when comparing pixel-wise differences normalized by $\sqrt{N}$, with $N$ the binned counts of a reference template; we stress this is an indicative diagnostic rather than a formal hypothesis test.

Undoubtedly, the \sky~diffuse templates can be used as an improvement over diffuse models based on Galactic gas maps and cosmic-ray propagation alone, which  typically leave residuals as large as tens of per cent even after optimization.

\paragraph{Injection and recovery tests.} Our second main interest is the sensitivity of \sky~to a DM signal in the GC. With the final goal of setting upper limits on the DM signal, we ask ourselves the question: In the case a DM signal is present in real data, would we wrongly exclude the corresponding mass and cross-section values?
Answering this question on synthetic datasets illustrates the capability of \sky~to recover the gamma-ray flux of a DM signal.
To this end, we designed the four different GCE composition scenarios discussed above, which we aim to recover via \sky. 

\begin{figure}[t]
     \includegraphics[width=\linewidth]{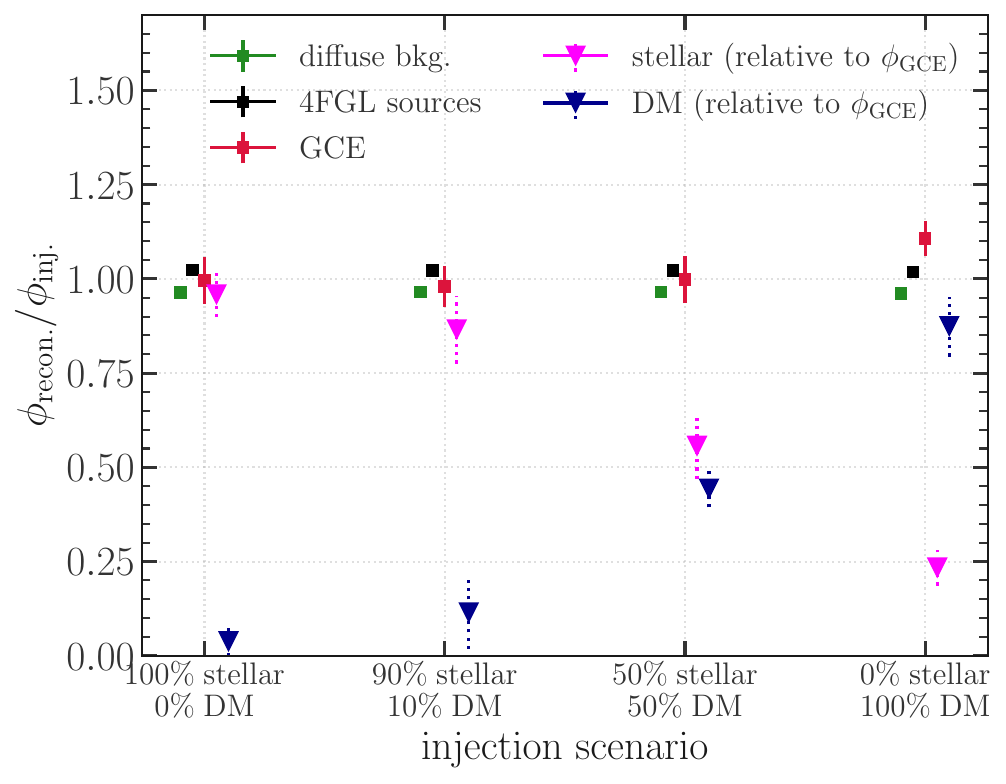}
     \caption{\textit{\sky~GCE emission injection and recovery test results.} Reconstructed flux $\phi_{\mathrm{recon.}}$ per injection scenario relative to the injected flux $\phi_{\mathrm{inj.}}$ for three main components: diffuse background (Galactic diffuse emission + DRGB; green), 4FGL catalog sources (black) and the GCE (red). In addition, we display the reconstructed flux of a DM (blue) and stellar bulge (pink) component relative to the injected total GCE flux $\phi_{\mathrm{GCE}}$. The flux values are integrated from 1.6 to 5.9 GeV. The quoted uncertainties were derived using eleven Poisson realizations of the underlying injection scenario. 
     }
      \label{fig:skyFACT-recovery}
\end{figure}

We summarize the results of the recovery tests in Fig.~\ref{fig:skyFACT-recovery}, comparing all four scenarios next to each other. The displayed error bars are derived from eleven Poisson realizations of the synthetic dataset. In this plot, we consider three main components: the diffuse background consisting of the Galactic diffuse emission components and DGRB, the 4FGL sources (point-like and extended) as well as the GCE (sum of stellar bulge and DM). We display their reconstructed flux in the energy range from 1.6 to 5.9 GeV within our $40^{\circ}\times40^{\circ}$ ROI (without a Galactic plane mask) relative to the injected flux in the same energy range. Hence, a value of 1 indicates perfect reconstruction.  In addition, we show the reconstructed flux of DM and stellar bulge component relative to the injected GCE flux $\phi_{\mathrm{GCE}}$. Per scenario, this value should correspond to the initial definition to achieve perfect recovery. 

Across the board, the diffuse and 4FGL sources are reconstructed identically, almost reaching 1 with negligible scatter. In fact, the 4FGL sources are reconstructed with a slightly larger flux than this component should exhibit. This observation can be naturally explained with the treatment of extended 4FGL sources within \sky. They are rather treated as a localized diffuse component with full spatial modulation freedom, so that it is very likely for them to absorb a part of the injected diffuse emission in the region where they are defined. We stress that the subsequent \onep~runs take the \sky~fit diffuse background plus extended sources as input. Overall, these two components are thus reliably and robustly recovered by the fit.
The total GCE flux is reconstructed around 1 at the $1\sigma$ level except for the scenario comprised of only DM, where we find more GCE emission than injected. Specifically:
\begin{itemize}
    
    \item Across the three injection scenarios with $<100\%$ DM, the total GCE flux is consistently reconstructed at the $1\sigma$ level. For the pure bulge case (100\% stellar bulge, 0\% DM), the injected emission is fully recovered, with the stellar component dominating and any spurious DM contribution remaining below 10\%. When a small DM fraction is injected (90\% bulge, 10\% DM), both components are correctly reconstructed within uncertainties, and the DM normalization is always detected at non-zero values, although the signal strength is close to the sensitivity limit of \sky. In the mixed case (50\% bulge, 50\% DM), both contributions are again accurately recovered. These results indicate that \sky~is capable of detecting a DM contribution down to the level of a few times below the thermal cross section, and could reliably recover a thermal-level signal in this mass range.
    
    \item 0\% stellar bulge, 100\% DM: We recover more GCE emission than injected around the $3\sigma$ level. A fraction of the emission comes from the Galactic diffuse emission and 4FGL sources, which lifts the not-injected stellar bulge component to a non-zero contribution in all realizations, averaging to about 25\% of $\phi_{\mathrm{GCE}}$. The DM component is recovered with about 90\% of its injected value on average. While the fluxes displayed in this plot are integrated from 1.6 to 5.9 GeV, inspecting the full recovered spectra reveals that the stellar bulge becomes relevant beyond 2 GeV, while it is suppressed below. The stellar bulge likely compensates for the deliberate spectral mismodeling of the \textit{Fermi} Bubbles in the synthetic data. A fact that would also explain the slightly reduced recovered flux of the diffuse component. It appears that the known degeneracy between the GCE and the low-latitude part of the \textit{Fermi} Bubbles becomes a crucial factor in this scenario. We stress, however, that in all realizations the normalization of the spectrum of DM gamma-ray emission is always higher than that of the stellar bulge in accordance with the prepared dataset.
\end{itemize}
The injection and recovery test results provide us with crucial information and a guideline to interpret the \sky~fit results we obtained from analyzing the real data: We observed that the normalization of the recovered DM component is a very good indicator for DM presence and, most of the times, \sky~correctly attributes the level of DM-related emission in the synthetic data to this component. However, as we demonstrated in the scenario of a GCE made solely from the stellar bulge, finding a non-zero DM template normalization does not automatically imply that a DM component is present in the data. In contrast, recovering a zero DM template normalization is a robust indicator that there is no DM admixture to the GCE or, at the very least, a small fraction below 10\%.
We remind that in real data we mostly find zero normalization of the DM component, except for a few points using the NFW126 profile that result in cross section values well below the upper limits we eventually derived in Sec.~\ref{sec:results-1ppdf}.
Our simulations make us confident that we are  not missing any possible DM signal in the \sky~fit.
On the other hand, it is noteworthy that the subsequent \onep~fit, despite initializing the diffuse background, DM, and stellar bulge templates to the best-fit \sky~results, maintains the freedom of re-fitting all components, therefore giving the possibility to suppress/enhance the DM signal and potentially alleviating \sky~mis-reconstruction. 
What is very relevant is that the mismodeling induced by the optimized diffuse model for the null hypothesis fit is within 2$\sigma$ (see Fig.~\ref{fig:skyFACT-mismodeling}. 
This level of residuals is likely the one faced by the \onep~while performing the subsequent real data fit. In the next subsection, we verify that this level of residuals would not alter the upper limits on DM.

Finally, we quantified the evidence for an additional DM component on top of the stellar bulge in a nested \sky~fit setting following the approach reported in Sec.~\ref{sec:SF_intro}. In the four tested scenarios of synthetic data, this metric does not always reliably extract the evidence of the injected DM component. As we did in all previous works on the GCE \cite{Calore:2021jvg, Manconi:2024tgh, Song:2024iup}, such evidence statements should be derived with additional means like a subsequent Bayesian template-based fit, as is the case for the \sky-\onep~hybrid method.

\subsection{Null hypothesis test}

\subsubsection{1pPDF simulation methods}

The \onep~simulations are designed to quantify if the obtained DM upper limits are robust against the null hypothesis, as well as the possible role of background systematics in shaping the upper limits. 
Our goal is to verify that the limits obtained within our methodology on real \fermi-LAT data are consistent with what we would obtain if we assume a simulated gamma-ray sky without a DM component, i.e., $A_{\rm DM}=0$. Additionally, we quantify the level of the systematic effects connected to the presence of mismodeling in the diffuse background components.  The performance of the \onep~method in reconstructing the $\mathrm{d}N/\mathrm{d}S$ of bright and faint point sources has been quantified in detail in \cite{Zechlin:1}. 

As we have done in \cite{Zechlin:1,Zechlin:2017uzo}, we use for this purpose realistic Monte Carlo simulations of \fermi-LAT gamma rays carried out using the \texttt{gtobssim} utility of the~\fermi~Science Tools. This allows us to include in the simulations not only the diffuse emissions and the detected sources, but also the full contribution from bright and faint point sources following a realistic $\mathrm{d}N/\mathrm{d}S$, which are reconstructed by the \onep~analysis, as well as realistic detector effects. 
A dedicated suite of 20 simulations covering the same time interval, event class and energy range matching the selection done for the real \fermi-LAT data is thus produced as described in what follows.  The mock events generated with \texttt{gtobssim} were then processed using the same analysis chain used to prepare the real data for the \onep~fit, including the computation of the effective PSF correction matching the simulation properties. 

Each simulated counts map contains two main components: point sources and diffuse emissions.
As for the point sources, their flux distribution is taken to follow the inner Galaxy $\mathrm{d}N/\mathrm{d}S$ observed in real data in the similar energy bin  from \cite{Calore:2021jvg}. 
Starting from this parametrization, the properties  of the point source list entering each \texttt{gtobssim} simulation are obtained from a dedicated Monte Carlo simulation which extracts flux values following a probability distribution obtained from the chosen $\mathrm{d}N/\mathrm{d}S$, as well as random, isotropic positions in the sky. Each simulation represents thus a different statistical realization of the input $\mathrm{d}N/\mathrm{d}S$. 
To each simulated source, we assign a flux spectrum following a power law, with photon index drawn from a Gaussian distribution of mean  $\Gamma=2.4$ and standard deviation $\sigma_{\Gamma}=0.2$. The minimum and maximum of the photon fluxes in the simulation are of $S_{\rm min}=10^{-13}$\,cm$^{-2}$\,s$^{-1}$ and $S_{\rm max}=10^{-8}$\,cm$^{-2}$\,s$^{-1}$, respectively. 
The diffuse emission model is taken to be the sum of the Galactic diffuse emission and the stellar bulge as found by the best \sky~fit of the real sky for a benchmark case, where also a DM component following a NFW profile with $\gamma=1.26$ and DM mass of $40$~GeV for the $b \bar{b}$ channel was included. The choice of the benchmark case is guided again by the fact that such mass and channel are usually found to be preferred to explain the full GCE with DM only.

The simulations are analyzed within the \onep~using a model including point sources, a Galactic diffuse emission, an isotropic background (accounting for sources too faint to be resolved even with the \onep), the stellar bulge, and additionally, a DM model (which is not included in the simulation) following the benchmark case quoted above. The mock data are analyzed in different ROIs covering the inner Galaxy, specifically testing different latitude cuts (from $0.5^\circ$ to $2^\circ$). The profile likelihood function for the normalization of the DM component $A_{\rm DM}$ is then extracted from the results of the 20 simulations, and compared to the real sky results to verify they follow similar statistical distributions. 

In the standard null hypothesis test, we use the same Galactic diffuse emission models to simulate and to analyze the results of the simulations within the \onep. To quantify possible effects coming from our non-perfect knowledge of the Galactic diffuse emission within the real sky, we repeat the null hypothesis test by analyzing the simulations with a different background model with respect to the one used to simulate the \fermi-LAT data. The goal is to see if the obtained limits for the DM component  change, i.e., to see if having diffuse emission mismodeling affects our sensitivity to a DM signal. For this, the model used to analyze the simulations is additionally processed by injecting a level of residual background mismodeling which follows what is quantified in the dedicated \sky~simulations (cf.~Sec.~\ref{app:skyfact-sim-results}).

\subsubsection{Results}

\begin{figure}[t]
\centering 
\includegraphics[width = 0.5\textwidth]{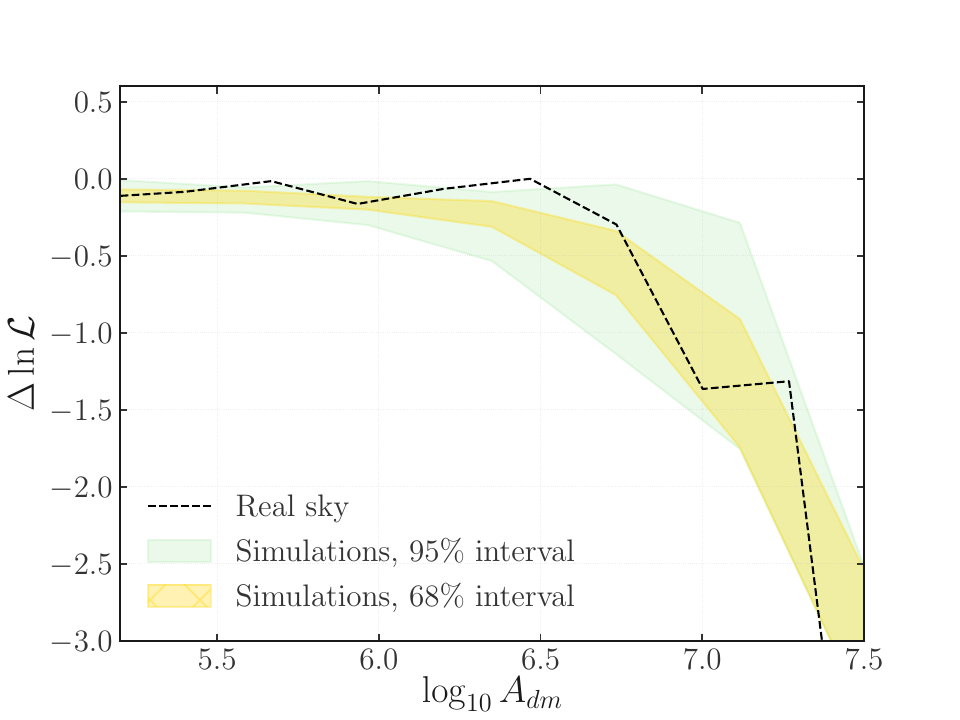}
\caption{\textit{Null hypothesis test for the \onep.} Comparison of the statistical expectation for the null hypothesis as derived by analyzing simulations with the results using real sky data, both using the \onep~ method.  
 The shaded bands depict the 68\% (yellow) and 95\% (green) confidence intervals derived from the statistical scatter of
the profile likelihood function for the normalization of the DM component, as obtained from simulations of the gamma-ray sky by assuming $A_{DM}$ = 0. The dashed
black line shows the corresponding result obtained from the actual data.}
\label{fig:null_hyp}
\end{figure}

Following the optimization process designed in \cite{Zechlin:2017uzo} for similar purposes, the suite of simulations is analyzed using the \onep~pipeline varying the defined ROI. For each tested ROI, we analyze the full suite of simulations. The statistical expectations for the null hypothesis are then extracted from the statistical scatter of the profile likelihood function for the normalization of the DM component $A_{\rm DM}$, by computing 68\% and 95\% confidence intervals. These statistical expectations, obtained by analyzing the null hypothesis simulations, where $A_{\rm DM}$ has been set equal to zero, are  compared with the results obtained analyzing the real sky data within the same ROI. 

We start with testing the ROI used to analyze a similar energy range within the inner Galaxy with the \onep~in \cite{Calore:2021jvg}, i.e. a 20 x 20 square degrees  centered at the GC, and varying the Galactic plane mask from 0.5 degrees to 2 degrees. We find that the cut at $1$ degrees provides the most stable results, and we thus select this ROI for extracting the DM constraints, as reported in the main text. 
The results for the selected ROI are reported in Fig.~\ref{fig:null_hyp}, where we show that the real sky profile likelihood for the $A_{\rm DM}$ parameter is perfectly compatible with the null hypothesis expectations at the 68\% confidence level, and its statistical fluctuations are included within the 95\% interval.  
The \dnds~parameters reconstructed by the \onep~when analyzing the simulations are found to be compatible with the injected source count distribution within the statistical errors. 

We have verified that reducing the latitude cut to $0.5$ degree provides stronger constraints, which are found to be marginally consistent with the lower 95\% interval. Conversely, increasing the latitude cut at $2$ degrees results in less constraining results, still compatible with the upper end of the 95\% interval. The choice of the $1$ degree cut is thus found to be a good trade-off between robustness and constraining power.

Fixing the ROI to the 20 $\times$ 20 degrees square with the latitude cut at $1$ degrees, we repeat the null hypothesis test using a diffuse emission model with an injected level of residual background mismodeling, following the methodology illustrated above. Also in this case, the full suite of null hypothesis simulations is analyzed using the \onep. 
When comparing to the statistical bands shown in Fig.~\ref{fig:null_hyp}, we find that the expected average of the upper limit on the DM normalization is shifted to larger values.  We interpret this as follows. Assuming that our optimized background model obtained by applying \sky~to the real data very likely does contain diffuse mismodeling, our upper limits end up being  weaker than they could be without such effects. In other terms, in principle, one could ''hide'' a larger DM signal in the data by a mismodeled background template. 
Nevertheless, the real sky results are still found to be compatible with the null hypothesis. Interestingly, for a couple of simulations, the \onep~results show a double-sided profile for the DM normalization, thus hinting at a DM detection, which is never the case when analyzing the real sky. If we were to see such a two-sided distribution in the real sky, we would likely be in a situation in which we cannot exclude the possibility that the evidence is coming from diffuse mismodeling. Since this does not happen in the real sky, we conclude that our tests demonstrate the robustness of the method against this source of systematics.

\bibliography{biblio} 

\end{document}